# Symmetry selective Dynamic Casimir Effect in One-Dimensional Photonic Crystals

Shaojie Ma[1, 2], Haixing Miao[2], Yuanjiang Xiang[1*], Shuang Zhang[2*]

[*] Corresponding authors: Y. Xiang: yjxiang@szu.edu.cn;

S. Zhang: S.Zhang@bham.ac.uk

*1Key Laboratory of Optoelectronic Devices and Systems of Ministry of Education and Guangdong Province, College of Optoelectronic Engineering, Shenzhen University, Shenzhen 518060, China*

*[2]School of Physics and Astronomy, University of Birmingham, Birmingham B15 2TT, United Kingdom*

**Abstract**: **Real photon pairs can be created in a dynamic cavity with periodically modulated refractive index of the constituent media or oscillating boundaries. This effect is called Dynamic Casimir effect (DCE), which represents one of the most amazing predictions of quantum field theory. Here, we investigate DCE in a dynamic one-dimensional photonic crystal system with both temporal and spatial modulation of the refractive index profile. Such a system can resonantly generate photons at driving frequencies equal to even or odd integer times of that of the fundamental cavity mode governed by the symmetry of the spatial modulation. We further observe interesting spectral and scaling behaviors for photons excited at the band edge. Our discovery introduces a new degree of freedom to enhance the efficiency of DCE.**

Dynamic Casimir effect [1-8] describes the creation of photon pairs from a quantum vacuum when the boundary conditions or the material properties are dynamically changed. Due to a time dependent geometric/material configuration, the quantum field experiences a parametric amplification processes over time, which is accompanied by generation of real quanta. Among various studied DCE systems, the most frequently discussed configurations are one-dimensional homogeneous cavities with vibrating boundaries [1,2] and cavities with fixed boundaries but dynamically changing optical properties [3,4]. Usually, a harmonically driven configuration [5-8] is employed to enhance DCE through resonant excitation at certain frequencies. In this process, DCE predicts a direct energy conversion from mechanical kinetic energy to electromagnetic energy. However, to observe a detectable signal, the moving speed of the geometrical configuration must be a substantial fraction of the speed of light, which becomes unrealistic for a macroscopic mechanical system [9]. Various solutions have been theoretically proposed to achieve rapid modulation of the cavity resonances, e.g., by using laser/magnetic field to modulate the reflection property of the semiconductor [10-11] / superconducting [12-13] mirror, or using modulation of nonlinear [14] materials, cold matter [15] material or superconductor circuits [16] to vary the refractive index of the filler inside the cavity, or nonadiabatic time modulation of the light-matter coupling strength [17-18] in cavity QED systems. Recently, by using superconducting quantum interference devices (SQUIDs), the effective length of a cavity was tuned by modulating its inductance through a time-varying magnetic flux at very high frequencies (>10 GHz), leading to observation of single-mode squeezing signals of DCE in the microwave regime [13, 16, 19-22], which verified the DCE excitation prediction. Thus far, most previous theories and experiments were based on dynamic cavities with homogeneous media, which are weakly excited even at resonance. In this work we show that inhomogeneously varying optical properties in a cavity containing 1D photonic crystals (PhCs) can lead to observation of interesting selection rules for the resonant excitation condition and greatly enhanced DCE photon generation rate for the edge modes governed by the symmetry of the spatial modulation.

Here, we look into one-dimensional photonic crystals with finite unit-cells and perfect electric conductor (PEC) boundaries, in which the permittivity of each slab is harmonically modulated (shown in Fig 1(a)). We discuss the squeezing (degenerate parametric process for producing single-mode squeezing) and acceleration (non-degenerate parametric process for producing two-mode squeezing) effect caused by the variation of resonance frequencies and eigenmodes, and numerically calculate the photon creation in these processes. For different spatial symmetries of the dynamic perturbations of the PhC systems, a significant DCE may occur only for resonant excitation at even or odd integer harmonics of the fundamental resonance frequency. The created photon intensity possesses interesting scaling behaviors with the number of unit cells. Our results provide a powerful degree of freedom to selectively enhance the rate of the DCE photon generation process.

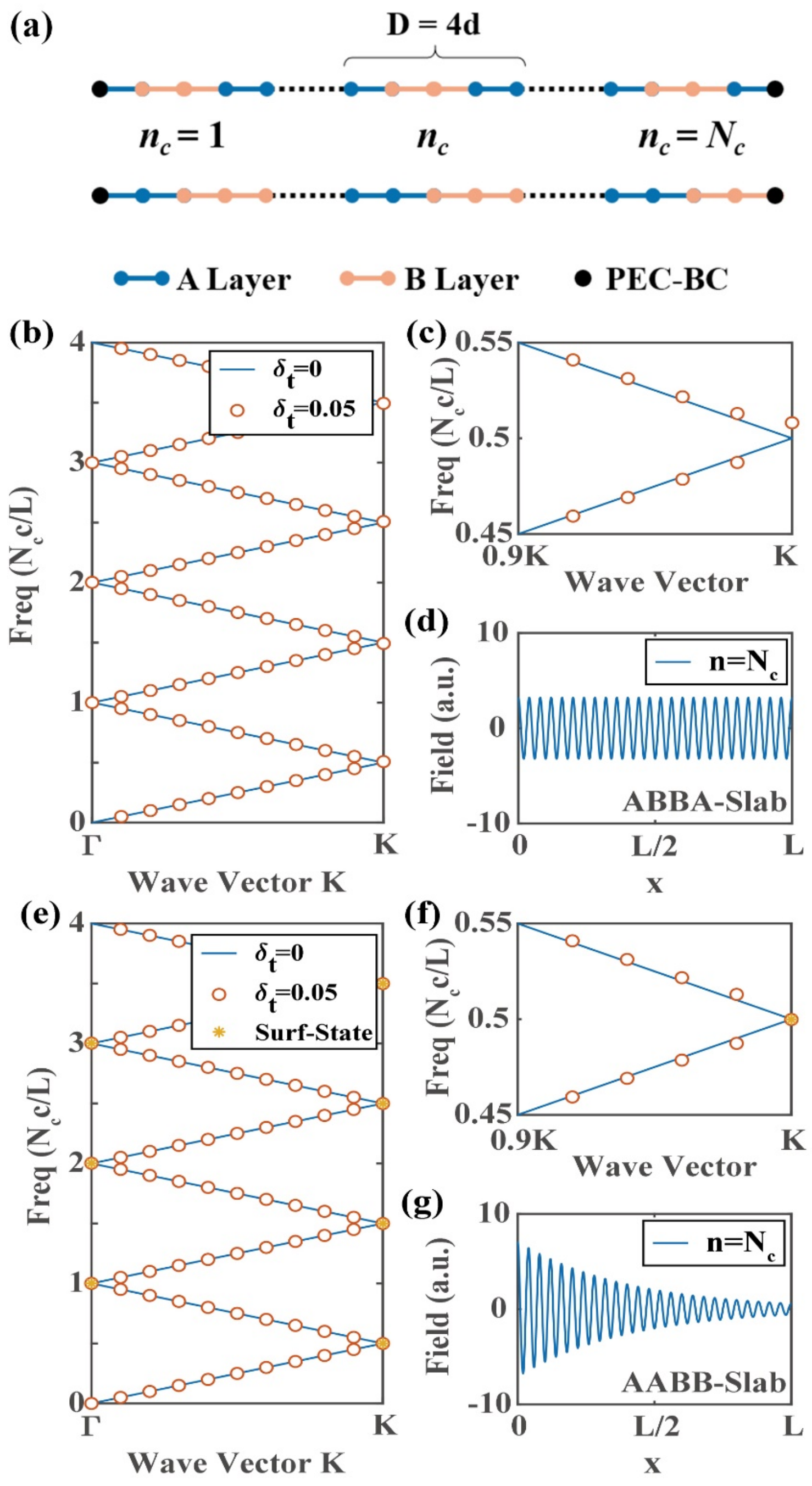

Figure 1 (a) Schematics of the a typical PhC system with ABBA and AABB

configuration. (b) and (e) are the band structures for these two systems with different permittivity variation, in which the number of repeated unit cells $N_c = 10$. (c) and (f) show the corresponding zoomed results near the first band edge denoted by the *K*-point with $N_c = 50$. (d) and (g) show the field distributions of the first *K*-point eigenmode.

As schematically depicted in Fig. 1(a), a PhC system consists of $N_c$ repeated unit cells inserted in a cavity formed by two PEC boundaries. Each unit consists of four layers with the same thickness $d$ but different dynamic varying permittivity $\varepsilon(t)$, marked as A and B with $\varepsilon_{A(B)}(t) = \varepsilon_0\,[1+(-)\,\delta_t]$, in which $\delta_t=\delta_0\cdot sin(\omega_d\cdot t)$ with $\delta_0$ and $\omega_d$ denoting the modulation amplitude and frequency, respectively. For simplicity, we only consider a small perturbation configuration with $\delta_0 << 1$. In this paper, we will focus on two typical setups - ABBA and AABB with spatially symmetrical and anti-symmetrical perturbation terms, respectively. For both systems, at an arbitrary time $t$, we can solve its instantaneous eigenmodes through Transfer Matrix Method (TMM) [23], in which both their eigenfrequency and eigenfunction vary harmonically in time. Fig. 1 (b-g) show the band structures and the typical field distributions of these two systems with $N_c$ repeated unit cells. Without the time dependent perturbation $\delta_0 = 0$, the system is just a homogeneous vacuum cavity with standing wave eigenmodes and its band structure is located at a series of equally separated points $(K_n, \omega_n) = (n_s\pi/L, n\pi c/L)$, in which $n_s$ denotes folded index of the eigenmode index $n$ to the 1$^{st}$ Brillouin zone, and $L=4N_cd$ denotes the effective optical length of the cavity. Clearly, there are $N_c$ eigenmodes in each band which equals to the number of repeated unit cells. With a small perturbation $\delta_t$, the band structure exhibits only a noticeable deformation close to the band edge denoted by the *K*-point due to degenerate perturbation. But for the two systems being investigated, completely different behaviors are observed near the *K*-point: the system with symmetric perturbation (ABBA system) has an obvious frequency shift, while that with antisymmetric perturbation (AABB system) exhibits a very slight frequency shift but the presence of an edge eigenmode, as shown in Fig. 1 (b-g). This will result in very different behaviors in photon creation process through the DCE in these two systems.

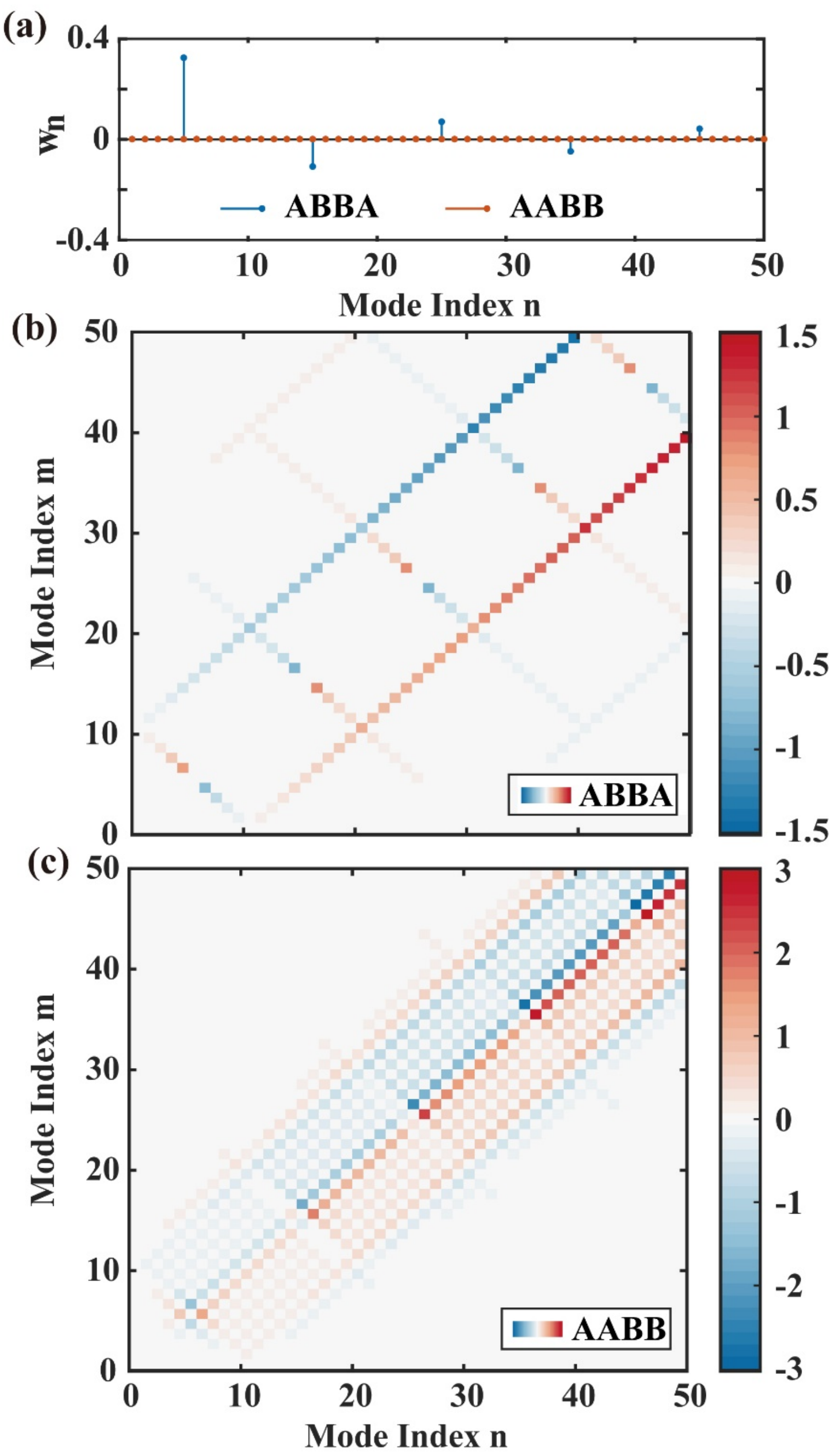

Figure 2 (a) Schematics of amplitude function $w_n$ of the squeezing term for each eigenmode in two systems. (b)-(c) Schematics of amplitude matrix distributions of the acceleration effect matrix $C_{mn}$ in two systems. In all these calculations, we choose $N_c = 5$ and solve them by using perturbation method.

Following the method given in Ref. [5,24-28], we can write down the evolution equation for the $n^{\text{th}}$ canonical annihilation operator $\hat{a}_n$ as:

$$\frac{\partial \hat{a}_n}{\partial t} = -i\omega_n \hat{a}_n + \xi_n \hat{a}_n^+ - \sum_m \left[ \mu_{mn}^+ \hat{a}_m + \mu_{mn}^- \hat{a}_m^+ \right] \ , \quad (1)$$

, with $\xi_n = \frac{1}{2} \cdot \frac{\partial \log \omega_n}{\partial t}$ , $\mu_{mn}^{\pm} = \frac{1}{2} \cdot \left( \sqrt{\frac{\omega_n}{\omega_m}} \pm \sqrt{\frac{\omega_m}{\omega_n}} \right) \cdot M_{mn}$ and $M_{mn} = \langle n | \frac{\partial}{\partial t} | m \rangle$ . On the RHS of Eq. (1), the first term is the conventional harmonic term, while the other two terms represent the two DCE contributions responsible for the creation of photon pairs: the squeezing term $\xi_n$ arising from the variation of the resonance frequency, and the acceleration term $\mu_{mn}^{\pm}$ caused mainly by the variation of the eigenfunction.

For a given harmonic perturbation parameter $\delta_t$, the two DCE contributions are expressed as:

$$\xi_n = \frac{1}{2} \cdot \frac{\partial \log \omega_n}{\partial \delta_t} \cdot \delta_0 \omega_d \cos(\omega_d t) \ \text{ and } \ M_{mn} = \langle n | \frac{\partial}{\partial \delta_t} | m \rangle \cdot \delta_0 \omega_d \cos(\omega_d t) \ , \quad (2)$$

which are both modulated at frequency $\omega_d$. For a small perturbation parameter $\delta_t$, the amplitude of these terms can be solved through a well-established perturbation framework [29-31] with the standing wave eigenmodes in a non-perturbed homogeneous cavity as the initial states [32]. The two amplitude functions can be analytically solved as:

$$w_n \equiv \frac{\partial \log \omega_n}{\partial \delta_t} = \frac{1}{2L} \cdot \int_0^L dx \cdot h(x) \cdot \left[ 1 - \cos\left( \frac{2n\pi}{L} x \right) \right] \quad (3)$$

$$C_{mn} \equiv \langle n | \frac{\partial}{\partial \delta_t} | m \rangle = \frac{mn \cdot (1 - \delta_{mn})}{(m^2 - n^2)} \cdot \frac{1}{L} \int_0^L dx \cdot h(x) \left\{ \cos\left[ (m-n) \frac{\pi}{L} x \right] - \cos\left[ (m+n) \frac{\pi}{L} x \right] \right\} \quad (4)$$

with $h(x) = \frac{\varepsilon_0}{\delta_t} \cdot \left( \frac{1}{\varepsilon(x)} - \frac{1}{\varepsilon_0} \right) = \pm \frac{\varepsilon_0}{\varepsilon(x)} \approx \pm 1$ and the sign "+/-" for "B/A" slab, respectively. This rectangular waveform distribution function $h(x)$ is determined by the specific PhC configuration and determines the selection rules in the DCE process.

In order to understand the influences of $h(x)$ in different PhC configurations, we analyze the concrete forms of the amplitude functions in the squeezing and acceleration terms. Clearly, from Eq. (3-4), both two amplitude functions are closely related to the Fourier components $h_n$ of the distribution function defined by $h_n = \int_0^L \varphi_n^* \cdot h(x) dx$, where $\varphi_n(x) = \sqrt{2/L} \cdot \cos(2n\pi \cdot x/L)$ is an even function defined in [0, $L$] region. It is evident that $h_n$ is strongly related to the symmetry of the perturbation.

In both ABBA and AABB configurations, the amplitude function $w_n$ in the squeezing term can be expressed as $w_n = -\sqrt{1/8L} \cdot h_n$. In the ABBA configuration with symmetric perturbation, an even $h(x)$ renders $h_n$ nonzero only when $n = n_K \equiv (2s+1) \cdot N_c$ ($s$ is an arbitrary integer), which corresponds to the modes at the edge of the Brillouin zone. This is consistent with the observed frequency shift at those $K$-points in the ABBA system. On the contrary, in the AABB configuration with antisymmetric perturbation, a nearly odd $h(x)$ function makes $h_n$ zero for all modes, which is consistent with the nearly $\delta_r$-independent eigenfrequency in the AABB system. The numerical results of $w_n$ presented in Fig. 2(a) for the two configurations with five repeated unit cells verify these perturbative analyses.

The amplitude function $C_{mn}$ of acceleration contribution is an antisymmetric sparse matrix which shows very distinct features for the symmetric and antisymmetric systems. Through some detailed analysis [32], it is shown that the matrix element is only nonzero when the index pair [$m$, $n$] satisfy $m \pm n = 2n_K$ for the symmetric system, or when they have different parities for the anti-symmetric system, as shown in Fig. 2 (b-c). Thus, the spatial symmetry of the system determines the distinct spectra and resonance conditions of photons creation process in these two systems.

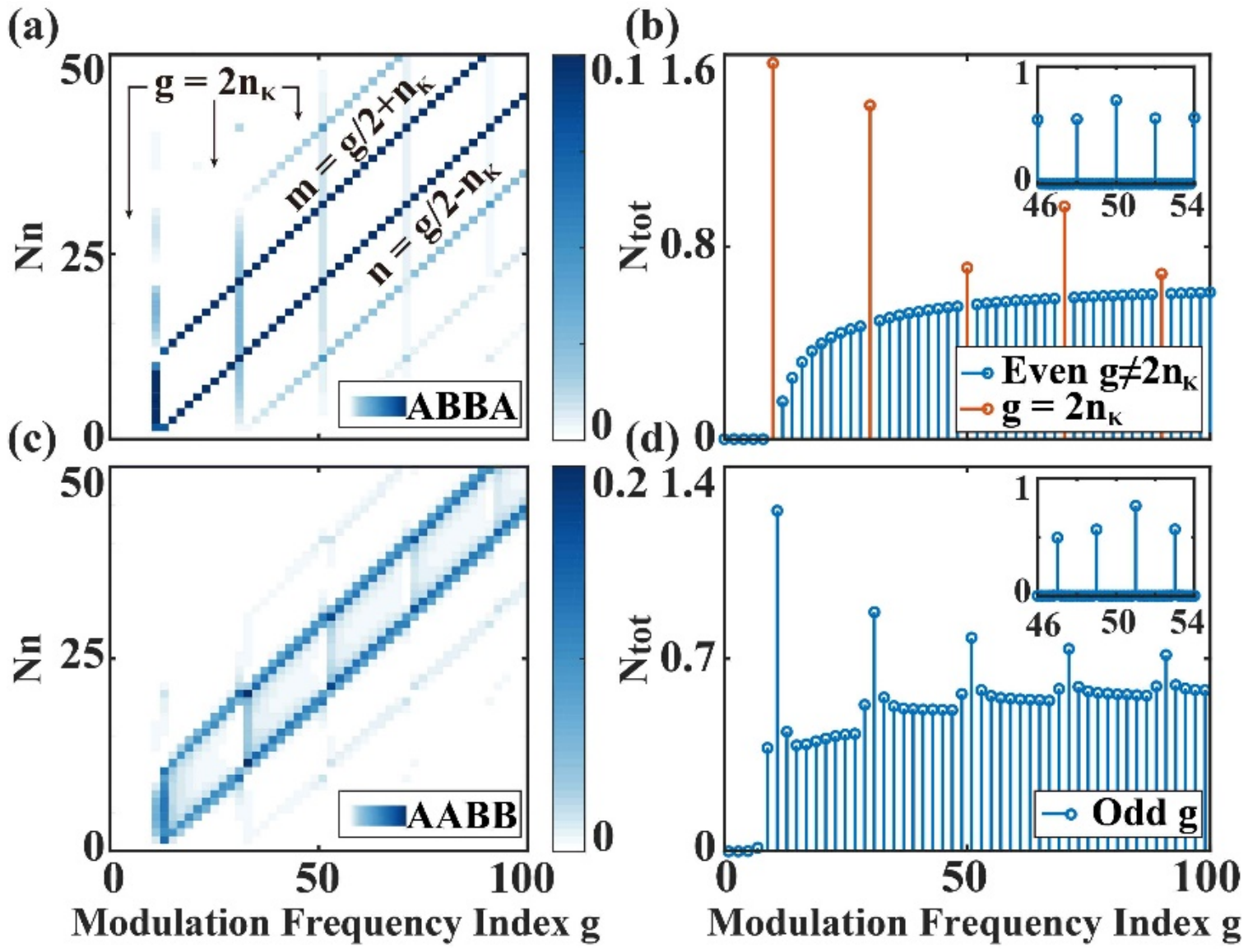

Figure 3 Schematics of the created photon spectra with different modulation

frequency $\omega_d$ at normalized time $\tau$=1 for (a) ABBA and (c) AABB system, respectively. (b) and (d) show the total number of created photon for even and odd g at the same time. Insets show the results for arbitrary g parameter. In all these calculations, we choose $N_c$= 5, $\delta_0$= 0.001 and solve them by using analytical method.

To further analyze the resonant conditions for the DCE process, we rewrite the evolution equation for the $n$-th canonical annihilation operator $\hat{a}_n$ in the interaction picture with $\bar{a}_n = \hat{a}_n \cdot \exp(i\omega_{n0}t)$, in which $\omega_{n,0} = n\pi c/L = n\omega_{1,0}$ is the $n^{\text{th}}$ resonance frequency of the non-perturbed homogenous cavity. In this way, both the squeezing/acceleration terms and the creation/annihilation operators contain fast oscillatory phase factor. For a significant resonance enhanced DCE effect, a harmonic excitation needs be applied to cancel out those time-varying factors. Neglecting the second derivatives and averaging over these fast oscillations [32, 33], we have:

$$\frac{\partial \bar{a}_n}{\partial \tau} = \frac{1}{4} w_n \cdot \delta\left(2\omega_{n,0} - \omega_d\right) \cdot \bar{a}_n^+ - \sum_m \begin{bmatrix} T_{mn}^+ \cdot \delta\left(\left|\omega_{n,0} - \omega_{m,0}\right| - \omega_d\right) \cdot \bar{a}_m \\ +T_{mn}^- \cdot \delta\left(\omega_{n,0} + \omega_{m,0} - \omega_d\right) \cdot \bar{a}_m^+ \end{bmatrix} \quad (5)$$

, in which $\tau = \delta_0 \omega_d t$ is the normalized time and $T_{mn}^{\pm} = \frac{1}{4} \cdot \left(\sqrt{n/m} \pm \sqrt{m/n}\right) \cdot C_{mn}$ is the amplitude of the acceleration effect. The RHS of Eq. (5) is independent with the strength of perturbation strength $\delta_0$.

For a given modulation frequency $\omega_d = g\pi c/L = g\omega_{1,0}$, where $\omega_{1,0}$ is the frequency of the fundamental non-perturbed cavity mode, a necessary condition to have contribution from the squeezing term $w_n$ is that $g$ is an even integer with $g=2n$, which leads to excitation of two $n$-mode photons. This, in conjunction with the preceding analysis of the amplitude function $w_n$ based on the parity of perturbation $h(x)$, leads to the conclusion that the photon pairs can be created by the squeezing effect only in the ABBA system with $g$= $2n_K$ for those K-points eigenmodes, while there are no excitations arising from the squeezing effect in the AABB system.

A similar argument is applied to the resonant excitation by the acceleration term $T_{mn}^{\pm}$. Two photons indexed $m$ and $n$ can be created/annihilated simultaneously for $g$=

$m+n$, whereas photons can be scattered from the $m$-th to the $n$-th mode or vice versa for $g=|m-n|$. In the ABBA system, since only those pairs satisfying $m\pm n=2n_k$ have nonzero $C_{mn}$ matrix elements, there exist two cases for the resonant excitation with an even $g$ parameter with $g\geq 2N_c$. In the first case with $g\neq 2n_K$, only specific photon pairs indexed $m=g/2+n_K$ and $n=g/2-n_K$ can be excited. While in the second case with $g=2n_k$, all mode pairs satisfying $g=m+n$ can be excited simultaneously. Meanwhile, in the AABB system, since only those pairs with $m\pm n=2s+1$ have nonzero $C_{mn}$ matrix elements, resonant excitation can occur for any odd integer value $g=2s+1$, in which many $[m, n]$ pairs with different parities can be excited simultaneously.

The excited photons spectra $N_n$ (number of excited photons in the $n$-th mode) and total numbers $N_{tot}$ (sum over all excited photons in all modes) over different modulation frequency $\omega_d$ are calculated by an analytical method [32], as shown in Fig. 3 for the two configurations consisting of 5 unit-cells at normalized time $\tau=1$. In the symmetrical ABBA system, only those with even $g$ parameters can resonantly create photon pairs. Interestingly, the excitations with $g=2n_K$ shows very different features from the rest, due to the contributions from both the squeezing term $w_n$ and the acceleration terms $T^{\pm}_{g\pm n,\, n}$, which results in a broadband $N_n$ spectrum in Fig. 3(a) and the red peaks in $N_{tot}$ spectra in Fig. 3(b). On the other hand, for $g\neq 2n_K$ the only contribution comes from the acceleration term $T^{-}_{g/2+n_K, g/2-n_K}$, which corresponds to a sparse $N_n$ spectrum and a lower value in $N_{tot}$. In the AABB system, only those odd $g$ parameters can resonantly create photon pairs, as shown in Fig. 3 (c-d). Some stronger peaks exist near the $K$-points, which is caused by the relatively stronger eigenmode variation near the Brillouin zone boundary. All these evolutions are consistent with the selective excitation conditions discussed above. (see a more detailed analysis in supplementary [32])

For all these three resonant excitation conditions with different g's, the total number of created photons exhibits a quadratic growth in a short-time scale, while an exponential growth in a long-time scale [32], as shown in Fig. S2 in the supplementary material. This evolution pattern is consistent with the law $N_{tot}(t)=\sinh^2(\beta\cdot\delta_0\omega_d t)$ in the homogeneous cavities [6,8,34], with $\beta$ a dimensionless constant to describe the rate of DCE Photon generation process that depends on the specific configuration of the system, which also implies the photon pairs are created through a DCE process.

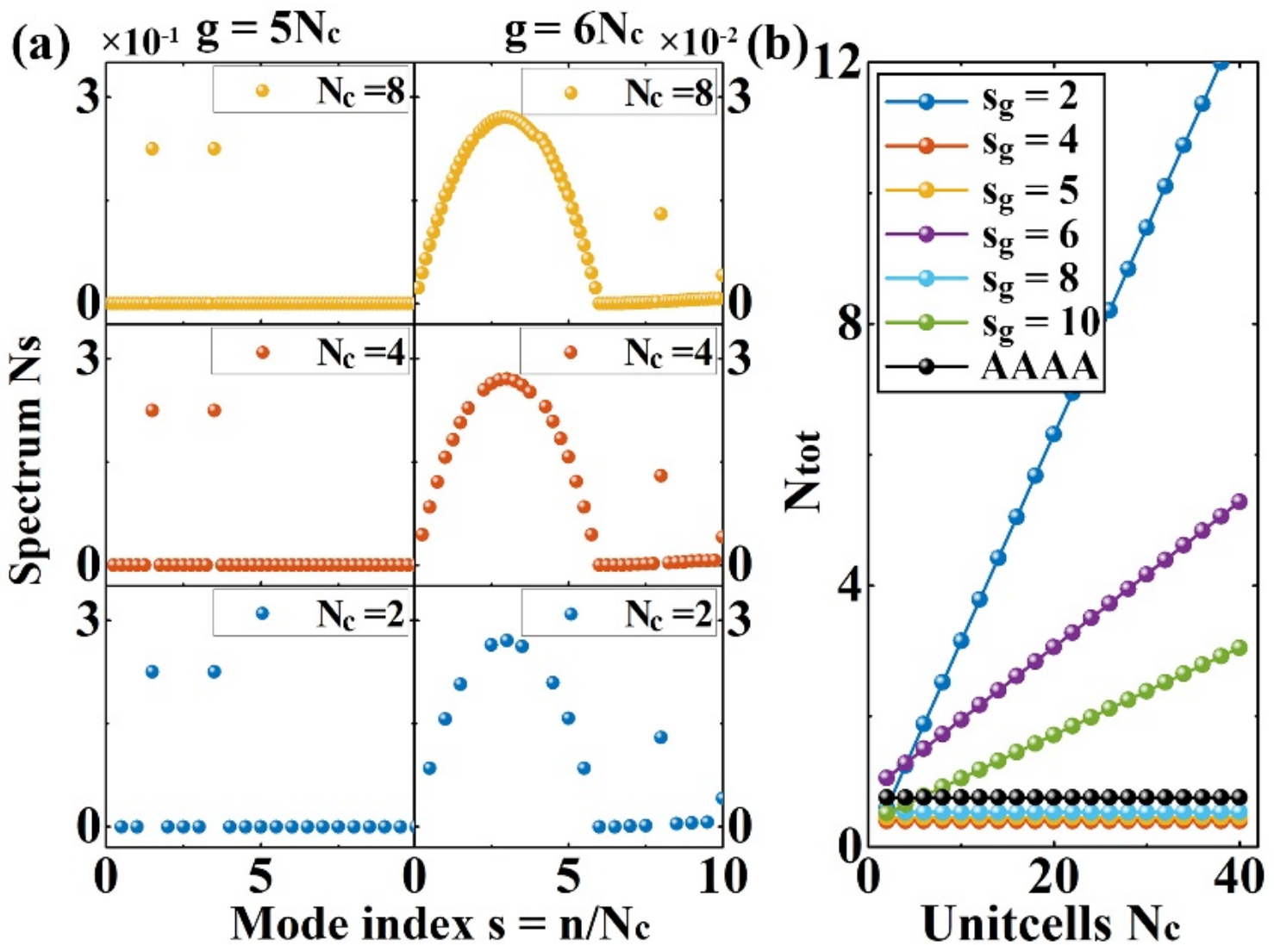

Figure 4 (a) Schematics of the excited photons spectra $N_s$ (number of excited photons in the ($s{\cdot}N_c$)-*th* mode) in ABBA system with different number of unit cells $N_c$ at normalized time $\tau$=1 with modulation frequency $\omega_d=5N_c{\cdot}\pi c/L$ ($s_g$=5) and $\omega_d=6N_c{\cdot}\pi c/L$ ($s_g$=6) respectively. (b) shows the total excited photons number $N_{tot}$ with different $N_c$ and $s_g$ parameters and in a homogeneously varying AAAA system with $s_g$=6. In all these calculations, we choose $\delta_0$= 0.001 and solve them by using analytical method.

For the symmetric ABBA system, we study the scaling behaviors of the number of created photons with the number of unit cells $N_c$, shown in Fig. 4. Here we renormalize both the mode index $n$ and modulation frequency index $g$ by $N_c$, with $n$= $s{\cdot}N_c$ and $g$= $s_g{\cdot}N_c$. Both the amplitude functions [$w_s$, $T^{\pm}_{s1,s2}$] and the modulation frequency remain a constant for different $N_c$ at fixed integer $s$ and $s_g$, which results in the same coefficients in Eq. (5). Therefore, the configurations with different numbers of unit cells have the same excitation spectra with respect to $s$, but different number of cavity modes per unit interval of $s$ parameter, as shown in Fig. 4 (a). However, for $g{\neq}2n_K$ only some discrete mode pairs are selectively excited, while for $g=2n_K$ all possible mode pairs are excited. Therefore, only those systems with modulation frequency corresponding to $g=2n_K$ can experience a strong enhancement of density of cavity modes with an increasing $N_c$. This explains the different scaling behaviors of the total number of excited photons $N_{tot}$ with the number of unit cells $N_c$, which exhibits a linear growth when $g=2n_K$ but remains a constant when $g\neq 2n_K$.

In order to have an intuitive understanding of the absolute DCE efficiency in PhCs, we compare our result with the conventional configuration of one-dimensional homogeneous cavity with dynamically changing optical properties, such as the setup in the SQUID experiments [13, 16], as shown in Fig. 4(b). When $N_c$ is small, our configuration has a similar efficiency with the conventional one, but it linearly increases for those modulation frequencies satisfying $g=2n_K$. Thus, our result offers a new degree of freedom to enhance the DCE signal in the experiment.

Moreover, in the supplementary material [32], we present a realistic ABBA PhC design based on a SQUID circuit that consists of a chain of coupled SQUID elements. We further numerically calculate the intensity of its output photon flux using the input-output theory, which clearly shows the superiority of our PhC cavity for enhancing the rate of the DCE photon generation process.

To summarize, we have studied DCE in a 1D photonic crystal system with dynamically modulated refractive indices in each layer, with the perturbation of index having symmetric or antisymmetric configuration. It was shown that the symmetry plays a very important role in selecting the DCE driving frequencies, and the corresponding photon generation exhibit intriguing scaling law for certain driving frequencies. Our result may inspire further experimental effort for achieving enhanced DCE in photonic systems.

S. M., Y. X. and S. Z. was supported by the European Research Council Consolidator Grant (TOPOLOGICAL), Horizon 2020 Action Project No. 734578 (D-SPA), the Engineering and Physical Sciences Research Council (EP/J018473/1). H. M. has been supported by UK STFC Ernest Rutherford Fellowship (Grant No. ST/M005844/11).

[1] G. T. Moore, “Quantum Theory of the Electromagnetic Field in a Variable-Length One-Dimensional Cavity”, Journal of Mathematical Physics **11**, 2679 (1970).

[2] C. K. Law, “Effective Hamiltonian for the radiation in a cavity with a moving mirror and a time-varying dielectric medium”, Physical Review A **49**, 433 (1994).

[3] V. V. Dodonov, A. B. Klimov, and D. E. Nikonov, “Quantum phenomena in nonstationary media”, Physical Review A **47**, 4422 (1993).

[4] H. Johnston and S. Sarkar, “Moving mirrors and time-varying dielectrics”, Physical Review A **51**, 4109 (1995).

[5] C. K. Law, “Resonance response of the quantum vacuum to an oscillating boundary”, Physical Review Letters **73**, 1931 (1994).

[6] V. V. Dodonov and A. B. Klimov, “Generation and detection of photons in a cavity with a resonantly oscillating boundary”, Physical Review A **53**, 2664 (1996).

[7] A. Lambrecht, M.-T. Jaekel, and S. Reynaud, “Motion Induced Radiation from a Vibrating Cavity”, Physical Review Letters **77**, 615 (1996).

[8] J. Y. Ji, H. H. Jung, J. W. Park, and K. S. Soh, “Production of photons by the parametric resonance in the dynamical Casimir effect”, Physical Review A **56**, 4440 (1997).

[9] V. V. Dodonov, “Current status of the dynamical Casimir effect”, Physica Scripta **82** (2010).

[10] M. Crocce, D. A. R. Dalvit, F. C. Lombardo, and F. D. Mazzitelli, “Model for resonant photon creation in a cavity with time-dependent conductivity”, Physical Review A 70 (2004).

[11] C. Braggio, G. Bressi, G. Carugno, C. D. Noce, G. Galeazzi, A. Lombardi, A. Palmieri, G. Ruoso, and D. Zanello, “A novel experimental approach for the detection of the dynamical Casimir effect”, Europhysics Letters **70**, 754 (2005).

[12] M. Tsindlekht, M. Golosovsky, H. Chayet, D. Davidov, and S. Chocron, “Frequency modulation of the superconducting parallel-plate microwave resonator by laser irradiation”, Applied Physics Letters **65**, 2875 (1994).

[13] C. M. Wilson, G. Johansson, A. Pourkabirian, M. Simoen, J. R. Johansson, T. Duty,

F. Nori, and P. Delsing, “Observation of the dynamical Casimir effect in a superconducting circuit”, Nature **479**, 376 (2011).

[14] V. V. Dodonov, “Dynamical Casimir effect in microwave cavities containing nonlinear crystals”, J. Phys.: Condens. Matter **27**, 214009 (2015).

[15] V. V. Dodonov and J. T. Mendonça, “Dynamical Casimir effect in ultra-cold matter with a time-dependent effective charge”, Physica Scripta T**160**, 014008 (2014).

[16] P. Lähteenmäki, G. S. Paraoanu, J. Hassel, and P. J. Hakonen, “Dynamical Casimir effect in a Josephson metamaterial”, Proceedings of the National Academy of Sciences **110**, 4234 (2013).

[17] S. De Liberato, C. Ciuti, and I. Carusotto, “Quantum Vacuum Radiation Spectra from a Semiconductor Microcavity with a Time-Modulated Vacuum Rabi Frequency”, Phys Rev Lett 98, 103602 (2007).

[18] M. Cirio, K. Debnath, N. Lambert, and F. Nori, “Amplified Optomechanical Transduction of Virtual Radiation Pressure”, Phys Rev Lett 119, 053601 (2017).

[19] J. R. Johansson, G. Johansson, C. M. Wilson, and F. Nori, “Dynamical Casimir effect in a superconducting coplanar waveguide”, Physical Review Letters **103**, 147003 (2009).

[20] C. M. Wilson, T. Duty, M. Sandberg, F. Persson, V. Shumeiko, and P. Delsing, “Photon generation in an electromagnetic cavity with a time-dependent boundary”, Physical Review Letters **105**, 233907 (2010).

[21] P. D. Nation, J. R. Johansson, M. P. Blencowe, and F. Nori, “Colloquium: Stimulating uncertainty: Amplifying the quantum vacuum with superconducting circuits”, Reviews of Modern Physics **84**, 1 (2012).

[22] C. Macklin, K. O’Brien, D. Hover, M. E. Schwartz, V. Bolkhovsky, X. Zhang, W. D. Oliver, and I. Siddiqi, “A near–quantum-limited Josephson traveling-wave parametric amplifier”, Science **350**, 307 (2015).

[23] R. Biswas, C. Chan, M. Sigalas, C. Soukoulis, and K. Ho, in *Photonic band gap materials* (Springer, 1996), pp. 23.

[24] R. Schützhold, G. Plunien, and G. Soff, “Trembling cavities in the canonical approach”, Physical Review A **57**, 2311 (1998).

[25] M. Ruser, “Vibrating cavities: a numerical approach”, Journal of Optics B: Quantum and Semiclassical Optics **7**, S100 (2005).
[26] M. Ruser, “Numerical approach to the dynamical Casimir effect”, Journal of Physics A: Mathematical and General **39**, 6711 (2006).
[27] M. Ruser, “Numerical investigation of photon creation in a three-dimensional resonantly vibrating cavity: Transverse electric modes”, Physical Review A **73**, 043811 (2006).
[28] P. I. Villar and A. Soba, “Adaptive numerical algorithms to simulate the dynamical Casimir effect in a closed cavity with different boundary conditions”, Physical Review E **96**, 013307 (2017).
[29] S. G. Johnson, M. Ibanescu, M. Skorobogatiy, O. Weisberg, J. Joannopoulos, and Y. Fink, “Perturbation theory for Maxwell’s equations with shifting material boundaries”, Physical review E **65**, 066611 (2002).
[30] S. G. Johnson, M. L. Povinelli, M. Soljačić, A. Karalis, S. Jacobs, and J. D. Joannopoulos, “Roughness losses and volume-current methods in photonic-crystal waveguides”, Applied Physics B **81**, 283 (2005).
[31] J. D. Joannopoulos, S. G. Johnson, J. N. Winn, and R. D. Meade, *Photonic crystals: molding the flow of light* (Princeton university press, 2011).
[32] See Supplemental Material for the rigorous/perturbation theory to solve the instantaneous eigenmodes, the detials to solve the photon excitation process in a closed/open system, the result in an asymmetric PhC system and the realitic design based on SQUID circuits.
[33] M. R. Setare, A. Seyedzahedi, “Fermion Particle Production as a Dynamical Casimir Effect inside a Three Dimensional Sphere”, Journal of Physics: Conference Series, **410**, 012150 (2013).
[34] V. V. Dodonov, A. B. Klimov, and D. E. Nikonov, “Quantum phenomena in resonators with moving walls”, Journal of mathematical physics **34**, 2742 (1993).

*Supplementary materials*

# Dynamic Casimir Effect in One-Dimensional Photonic Crystal

Shaojie Ma[1, 2], Haixing Miao[2], Yuanjiang Xiang[1,*] and Shuang Zhang[2, *]

[*] Corresponding authors: Y. Xiang: yjxiang@szu.edu.cn;

S. Zhang: S.Zhang@bham.ac.uk

[1]*Key Laboratory of Optoelectronic Devices and Systems of Ministry of Education and Guangdong Province, College of Optoelectronic Engineering, Shenzhen University, Shenzhen 518060, China*

[2]*School of Physics and Astronomy, University of Birmingham, Birmingham B15 2TT, United Kingdom*

# 1. Rigorous method to solve eigenmodes

In this section, we describe how to use the transfer matrix method (TMM) to solve the eigenmodes in the one-dimensional PhC system described in the main text. Our 1D system consists of periodic arrangement of homogeneous slabs. Within the $l$-th slab the field can be expressed as:

$$\begin{cases} E_{//}^{l}\left(x\right)=A_{l}e^{ik_{l}x}+B_{l}e^{-ik_{l}x} \\ H_{//}^{l}\left(x\right)=\dfrac{1}{Z_{l}}\left(A_{l}e^{ik_{l}x}-B_{l}e^{-ik_{l}x}\right) \end{cases}, \tag{S1.1}$$

in which $A_l$ and $B_l$ are the amplitudes of the forward and backward wave component, $k_l$ is the wave vector along x-direction, and $Z_l$ is the impedance of the $l$-th slab given by $Z_l=\sqrt{\mu/\varepsilon_l}$ . At the interface between adjacent layers the boundary conditions $E_{//}^{l}=E_{//}^{l+1}$ and $H_{//}^{l}=H_{//}^{l+1}$ need be satisfied. Thus, we can define a transfer matrix $M$ to relate the coefficients in adjacent slabs, with

$$\begin{pmatrix} A_{l+1} \\ B_{l+1} \end{pmatrix}=M_{l+1,l}\begin{pmatrix} A_{l} \\ B_{l} \end{pmatrix}=\frac{1}{2}\begin{pmatrix} 1+Z_{l+1}/Z_{l} & 1-Z_{l+1}/Z_{l} \\ 1-Z_{l+1}/Z_{l} & 1+Z_{l+1}/Z_{l} \end{pmatrix}\begin{pmatrix} A_{l} \\ B_{l} \end{pmatrix}, \tag{S1.2}$$

In the $l$-th slab, the forward and backward propagation fields on two boundaries separated by $d_l$ can be related to each other through a propagating matrix $P$, with:

$$P_{l}\left(d_{l}\right)=\begin{pmatrix} e^{ik_{l}d_{l}} & 0 \\ 0 & e^{-ik_{l}d_{l}} \end{pmatrix}. \tag{S1.3}$$

In our system, the PhC system has $N_c$ repeated unit-cells and two PEC boundaries. For each unit-cell, we define a $Q$ matrix to describe its scattering property. For the two setups ABBA and AABB, the $Q$ matrices are:

$$Q_{ABBA}=M_{AA}P_{A}M_{AB}P_{B}M_{BB}P_{B}M_{BA}P_{A} \;\;\&\;\; Q_{AABB}=M_{AB}P_{B}M_{BB}P_{B}M_{BA}P_{A}M_{AA}P_{A}, \tag{S1.4}$$

which are functions of angular frequency $\omega$.

At the PEC boundaries of the cavity at $x=0$ and $x=L$ ($L$ is the total length of the PhC system), the forward and backward propagation fields must add up to zero. The fields at the two boundaries are related by the overall transfer matrix of the cavity:

$$a\begin{pmatrix}1\\-1\end{pmatrix}=T\cdot\begin{pmatrix}1\\-1\end{pmatrix}=M_{last}{}^{-1}\cdot Q^{N_c}\cdot\begin{pmatrix}1\\-1\end{pmatrix}, \quad \text{(S1.5)}$$

in which $M_{\text{last}}= M_{AA}$ for ABBA system and $M_{\text{last}}= M_{AB}$ for AABB system, $a$ is the ratio between the field strengths on the two boundaries. In equation (S1.5), the overall transfer matrix $T$ should satisfy:

$$T_{11}-T_{12}+T_{21}-T_{22}=0, \quad \text{(S1.6)}$$

with $T_{ij} = T(i, j)$ being the $i$-th row, $j$-th column element of the transfer $T$ matrix. By solving Eq. S.1.6, one can obtain the instantaneous eigenmodes for each configuration. The results for the two configurations (ABBA and AABB) are shown in Fig. 1 in the main manuscript.

Here the Bloch wave vector $K_n$ of each eigenmode can be obtained by diagonalizing the $Q$ matrix of the $n$-th cavity mode with $Q_n = U\Lambda_n U^{-1}$. Considering the eigenvalue as the propagation phase of the wave vector, we have:

$$\Lambda_n=\begin{pmatrix}\exp(iK_nD) & \\ & \exp(-iK_nD)\end{pmatrix}\Rightarrow K_n=D^{-1}\cdot arc\cos\left[\frac{1}{2}Tr(Q_n)\right], \quad \text{(S1.7)}$$

with $D$ being the length of each unit cell. With a small perturbation amplitude $\delta_0$, the wave vector $K_n$ is very close to that in a vacuum cavity, with $K_n = n_s\pi/L + O(\delta_0{}^2)$.

# 2. Perturbation method to solve eigenmodes

In this section, we describe how to solve the eigenmodes in the inhomogeneous cavity. We also provide details on the two driving terms $\xi_n$ and $M_{mn}$ in the PhC system with time-independent perturbation theory [1-3].

For the instantaneous eigenmode in the 1D-PhC system at any moment $t$, the field should satisfy the wave equation:

$$\nabla\times\left[\frac{1}{\varepsilon_r(x)}\nabla\times H(x)\right]=\left(\frac{\omega}{c}\right)^2\cdot H(x), \tag{S2.1}$$

with $\varepsilon_r(x)$ being the instantaneous relative permittivity distribution. It can be rewritten in the form of a time-independent Schrödinger equation $\Theta\cdot\varphi=E\varphi$ with:

$$\Theta=\nabla\times\left[\frac{1}{\varepsilon_r(x)}\nabla\times\right],\quad \varphi=H(x),\quad E=\left(\frac{\omega}{c}\right)^2. \tag{S2.2}$$

In a non-perturbed configuration with $\delta_0=0$, the system becomes a homogeneous cavity with standing wave eigenmodes described by:

$$\Theta_0=\nabla\times(\nabla\times),\quad \varphi_{n0}=\sqrt{\frac{2}{L}}\cdot\cos\left(\frac{n\pi}{L}x\right)e_y,\quad E_n^{(0)}=\left(\frac{n\pi}{L}\right)^2. \tag{S2.3}$$

The eigen modes satisfy the orthogonality condition $\langle m|n\rangle=\int_0^L dx\varphi_{m0}^*\cdot\varphi_{n0}=\delta_{mn}$.

By using the standard perturbation method, we can obtain the first order correction term of the instantaneous eigenfrequencies and eigenmodes:

$$E_n^{(1)}=\langle n|\Theta_1|n\rangle,\quad \alpha_{nm}^{(1)}=\frac{\langle m|\Theta_1|n\rangle}{E_n^{(0)}-E_m^{(0)}}, \tag{S2.4}$$

with $E_n=E_n^{(0)}+E_n^{(1)}+\ldots$, $\varphi_n=\varphi_{n0}+\alpha_{nm}^{(1)}\varphi_{m0}+\ldots$ and

$$\Theta_1=\Theta-\Theta_0=\nabla\times\left\{\frac{1}{\delta_t}\left[\frac{1}{\varepsilon_r(x)}-1\right]\cdot\nabla\times\right\}\cdot\delta_t, \tag{S2.5}$$

is the perturbation term of the effective Hamiltonian with a small amplitude $\delta_t$.

Substituting the above perturbation term into Eq. (S2.4), we have:

$$\begin{aligned}\Theta_1|n\rangle&=\nabla\times\left\{\left[\frac{1-\varepsilon_r(x)}{\varepsilon_r(x)\delta_t}\right]\cdot\left[-\sqrt{\frac{2}{L}}\cdot\frac{n\pi}{L}\cdot\sin\left(\frac{n\pi}{L}x\right)e_z\right]\right\}\cdot\delta_t\\&=\frac{d}{dx}\left[\frac{1-\varepsilon_r(x)}{\varepsilon_r(x)\delta_t}\right]\sqrt{\frac{2}{L}}\cdot\frac{n\pi}{L}\cdot\sin\left(\frac{n\pi}{L}x\right)e_y\cdot\delta_t+\frac{1-\varepsilon_r(x)}{\varepsilon_r(x)\delta_t}\cdot\sqrt{\frac{2}{L}}\cdot\left(\frac{n\pi}{L}\right)^2\cos\left(\frac{n\pi}{L}x\right)e_y\cdot\delta_t\end{aligned} \tag{S2.6}$$

and therefore, $E_n^{(1)}$ and $\alpha_{nm}^{(1)}$ can be expressed as:

$$E_n^{(1)}=\langle n|\Theta_1|n\rangle=\delta_t\cdot\frac{2n^2\pi^2}{L^3}\int_0^L dx\frac{1-\varepsilon_r(x)}{\varepsilon_r(x)\delta_t}\sin^2\left(\frac{n\pi}{L}x\right) \tag{S2.7}$$

$$\alpha_{nm}{}^{(1)} = \frac{\langle m|\Theta_1|n\rangle}{E_n{}^{(0)} - E_m{}^{(0)}} = \delta_t \cdot \frac{2nm}{\left(n^2 - m^2\right)\cdot L}\int_0^L dx \frac{1-\varepsilon_r(x)}{\varepsilon_r(x)\delta_t}\sin\left(\frac{n\pi}{L}x\right)\sin\left(\frac{m\pi}{L}x\right) \tag{S2.8}$$

with a small perturbation amplitude $\delta_t$.

The core amplitude function of the squeezing term is:

$$\begin{aligned} w_n &= \frac{\partial\omega_n}{\omega_n\partial\delta_t} = \frac{1}{2}\cdot\frac{\partial E_n}{E_n\partial\delta_t} = \frac{E_n{}^{(1)}}{2E_n{}^{(0)}\cdot\delta_t} \\ &= \frac{1}{L}\int_0^L dx\frac{1-\varepsilon_r(x)}{\varepsilon_r(x)\cdot\delta_t}\sin^2\left(\frac{n\pi}{L}x\right) \\ &= \frac{1}{2L}\int_0^L dx\frac{1-\varepsilon_r(x)}{\varepsilon_r(x)\cdot\delta_t}\cdot\left[1-\cos\left(\frac{2n\pi}{L}x\right)\right] \end{aligned} . \tag{S2.9}$$

The core amplitude function of the acceleration term satisfies:

$$C_{mn} = \langle n|\partial/\partial\delta_t|m\rangle = \int_0^L dx\left[\phi_{n0}(x) + \alpha_{ns}^{(1)}\phi_{s0}(x) + ...\right]^* \cdot \left[\alpha_{mq}^{(1)}\phi_{q0}(x)\delta_t^{-1} + ...\right] = \alpha_{mn}{}^{(1)}\delta_t^{-1}, \tag{S2.10}$$

which can be further derived as:

$$\begin{aligned} C_{mn} &= \frac{2mn}{\left(m^2-n^2\right)\cdot L}\int_0^L dx\frac{1-\varepsilon_r(x)}{\varepsilon_r(x)\delta_t}\sin\left(\frac{m\pi}{L}x\right)\sin\left(\frac{n\pi}{L}x\right) \\ &= \frac{mn}{\left(m^2-n^2\right)\cdot L}\int_0^L dx\cdot\frac{1-\varepsilon_r(x)}{\varepsilon_r(x)\delta_t}\left\{\cos\left[(m-n)\frac{\pi}{L}x\right] - \cos\left[(m+n)\frac{\pi}{L}x\right]\right\} \end{aligned} . \tag{S2.11}$$

In both amplitude functions, the key term is the distribution function:

$$h(x) = \frac{\varepsilon_0}{\delta_t}\cdot\left(\frac{1}{\varepsilon(x)} - \frac{1}{\varepsilon_0}\right) = \frac{1-\varepsilon_r(x)}{\varepsilon_r(x)\delta_t} = \pm\frac{\varepsilon_0}{\varepsilon(x)} \approx \pm 1, \tag{S2.12}$$

with the sign "-" for "A" slab and "+" for "B" slab, which is determined by the slab distributions, instead of the perturbation amplitude $\delta_t$. The Fourier components of this distribution function will fully determine the selectivity of the squeezing/ acceleration terms in Eqs. (S2.9) and (S2.11).

To verify its accuracy, we compare the amplitude functions obtained by perturbation method and numerical method, and their results perfectly match for both PhC systems with different symmetry, as shown in Fig. S1.

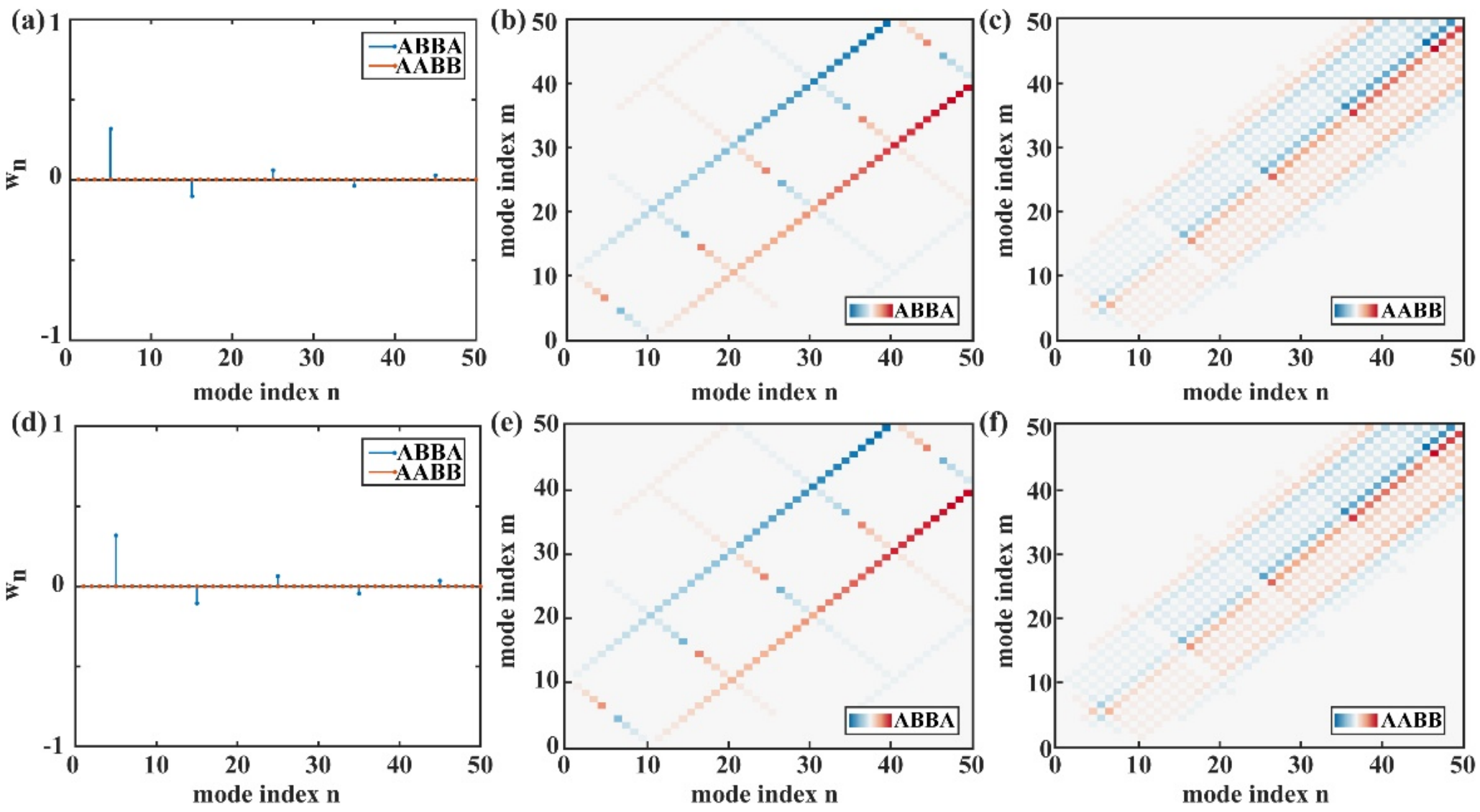

Figure S3. (a) Schematics of the amplitude function of the squeezing term for each eigenmode by numerical method. (b-c) Schematics of the amplitude matrix distributions of the acceleration effect for the (b) ABBA and (c) AABB systems by numerical method. (d-f) are the corresponding results of (a-c) obtained by the perturbation method. In all of these calculations, we choose $N_c$=5.

Clearly, the symmetry of PhC configuration can induce different sparse matrix $w_n$ and $C_{mn}$. In the main manuscript, the concrete forms of $h_n$ and $w_n$ have been discussed, which show very distinct features for the symmetric and anti-symmetric systems. The amplitude function $C_{mn}$ of the acceleration contribution is an anti-symmetric real matrix with $C_{mn} = -C_{nm} = \mathrm{Re}(C_{mn})$. From Eq. (S2.9) and (S2.11), for $n \pm m = 2l$ ($n$ and $m$ are both even or odd), $C_{mn}$ has nonzero value only when the Fourier component $h_l$ is nonzero. Since $h_l$ is nonzero only when $l=n_K$ in the ABBA system, only the eigenmode pairs satisfying $n \pm m = 2n_K$ can couple to each other. These eigenmode pairs have the same wave vector and nonzero $C_{mn}$ elements. On the other hand, for $n \pm m = 2l+1$, $C_{mn}$ connects two eigenmodes with different parities, therefore $C_{mn}$ is always zero in a symmetric system, whereas it has no restriction in an anti-symmetric system. Based on the discussion above, we can obtain all the acceleration matrix elements for the two systems, as shown in Fig. 2 (b-c). They are both sparse matrixes with nonzero elements only when the index pair [$m$, $n$] has same/different parity in symmetric/anti-symmetric configuration. This feature leads to the distinct spectra and resonance conditions for photon creation process in these two systems.

# 3. Derivation of Evolution equations for photon excitation process

In a 1D-PhC system, the field at any time can be projected into a set of instantaneous eigenmodes $\varphi_n(x;\, t)$ by canonical variables $q_n(t)$ as:

$$\varphi(x;t) = \sum\nolimits_n q_n(t)\varphi_n(x;t), \tag{S3.1}$$

in which $\varphi_n(x;t) = H_n(x;t)$ satisfy the wave equation:

$$\frac{\partial}{\partial x}\left(\frac{1}{\varepsilon_r(x;t)}\cdot\frac{\partial}{\partial x}\right)\varphi_n(x;t) = -\left[\frac{\omega_n(t)}{c}\right]^2 \varphi_n(x;t) \tag{S3.2}$$

The eigenmodes of the above equation satisfy the orthonormal relation:

$$\langle n|m\rangle = \int_0^L dx\varphi_n^*(x;t)\varphi_m(x;t) = \delta_{nm}, \tag{S3.3}$$

in which $L$ is the length of the PhC cavity.

A general field $\varphi(x;\, t)$ satisfies the wave equation:

$$\frac{\partial}{\partial x}\left(\frac{1}{\varepsilon_r(x;t)}\cdot\frac{\partial}{\partial x}\right)\varphi(x,t) = \frac{\partial^2}{c^2\partial t^2}\varphi(x,t) \tag{S3.4}$$

Inserting Eq. (S3.1) into (S3.4), multiplying it by $\varphi_n^*(x;\, t)$ and integrating over $L$ leads to the equations of motion for the canonical variables $q_n(t)$:

$$q_n(t)+\omega_n(t)^2\cdot q_n(t) + 2\sum\nolimits_m\langle n|m\rangle\cdot q_m(t) + \sum\nolimits_m\langle n|m\rangle\cdot q_m(t) = 0 \tag{S3.5}$$

Following the traditional dynamic Casimir effect calculation process, we define two matrices:

$$\begin{cases} M_{mn} = \langle n|m\rangle \\ N_{mn} = \langle n|m\rangle = M_{mn} - \langle n|m\rangle \end{cases}, \tag{S3.6}$$

in which $M$ is an anti-symmetric matrix with $M_{mn} = -M_{nm}$. Besides, $M_{nn} = 0$ due to the constant inner product of any two modes: $\langle n|m\rangle = \delta_{nm}$. Therefore, the motion equation for the canonical variables $q_n(t)$ is:

$$q_n(t)+\omega_n(t)^2\cdot q_n(t)+2\sum\nolimits_m M_{mn}\cdot q_m(t)+\sum\nolimits_m\left(M_{mn}-N_{mn}\right)\cdot q_m(t)=0, \quad \text{(S3.7)}$$

which is just the same as that for the canonical variables $q_n(t)$ in a cavity with homogeneous media and vibrating mirrors. The only difference is the definition of resonance frequency $\omega_n(t)$ and eigenfunction $\varphi_n(x;\, t)$. Following the derivation of the motion equation for the canonical annihilation/creation operator, we can obtain the evolution equations Eq. (1) and (5) in the main text, which are just the same as the traditional one in a homogeneous cavity.

Defining the momentum operators as $p_n(t)=q_n(t)+\sum\nolimits_n M_{mn}q_m$, the motion equation (S3.6) can be replaced by two coupled first-order differential equations:

$$\begin{cases} q_n = p_n - \sum\nolimits_m M_{mn}q_m \\ p_n = -\omega_n(t)^2 q_n + \sum\nolimits_m M_{nm}p_m \end{cases} \quad \text{(S3.8)}$$

Because these equations are linear with respect to "$q$" and "$p$", they are formally invariant under the standard canonical quantization procedure which replaces the Poisson bracket by the quantum commutator. Therefore, Eq. (S3.8) can be readily elevated to be the quantum Heisenberg equation of motion for the canonical variables "$q$" and "$p$", which are now the quantum operators and satisfy the general equal-time commutation relations for position and momentum operators. [4].

In order to describe the system in Fock space, we introduce the instantaneous creation and annihilation operators:

$$\begin{cases} a_n^{+} = \dfrac{1}{\sqrt{2\omega_n(t)}}\left[\omega_n(t)q_n(t)-ip_n(t)\right] \\ a_n = \dfrac{1}{\sqrt{2\omega_n(t)}}\left[\omega_n(t)q_n(t)+ip_n(t)\right] \end{cases} \quad \text{(S3.9)}$$

Because of the explicit time dependence of eigenfrequency $\omega_n(t)$ and the commutation relations for position and momentum operators, the evolution equation of these operators has some extra terms, which is:

$$\begin{cases} \dfrac{\partial \hat{a}_n}{\partial t} = -i\omega_n \hat{a}_n + \dfrac{1}{2}\cdot\dfrac{\partial \omega_n}{\omega_n \partial t}\hat{a}_n^+ - \dfrac{1}{2}\sum_m \left[ \left( \sqrt{\dfrac{\omega_n}{\omega_m}} + \sqrt{\dfrac{\omega_m}{\omega_n}} \right)\cdot M_{mn}\hat{a}_m + \left( \sqrt{\dfrac{\omega_n}{\omega_m}} - \sqrt{\dfrac{\omega_m}{\omega_n}} \right)\cdot M_{mn}\hat{a}_m^+ \right] \\ \dfrac{\partial \hat{a}_n^+}{\partial t} = +i\omega_n \hat{a}_n^+ + \dfrac{1}{2}\cdot\dfrac{\partial \omega_n}{\omega_n \partial t}\hat{a}_n - \dfrac{1}{2}\sum_m \left[ \left( \sqrt{\dfrac{\omega_n}{\omega_m}} + \sqrt{\dfrac{\omega_m}{\omega_n}} \right)\cdot M_{mn}\hat{a}_m^+ + \left( \sqrt{\dfrac{\omega_n}{\omega_m}} - \sqrt{\dfrac{\omega_m}{\omega_n}} \right)\cdot M_{mn}\hat{a}_m \right] \end{cases}$$

(S3.10)

It can be abbreviated as:

$$\begin{cases} \dfrac{\partial \hat{a}_n}{\partial t} = -i\omega_n \hat{a}_n + \xi_n \hat{a}_n^+ - \sum_m \left[ \mu_{mn}^+ \hat{a}_m + \mu_{mn}^- \hat{a}_m^+ \right] \\ \dfrac{\partial \hat{a}_n^+}{\partial t} = +i\omega_n \hat{a}_n^+ + \xi_n \hat{a}_n - \sum_m \left[ \mu_{mn}^+ \hat{a}_m^+ + \mu_{mn}^- \hat{a}_m \right] \end{cases}, \quad \text{(S3.11)}$$

when we define two driving terms:

$$\xi_n = \frac{1}{2}\cdot\frac{\partial \omega_n}{\omega_n \partial t}, \mu_{mn}^{\pm} = \frac{1}{2}\cdot\left( \sqrt{\frac{\omega_n}{\omega_m}} \pm \sqrt{\frac{\omega_m}{\omega_n}} \right)\cdot M_{mn}. \quad \text{(S3.12)}$$

Here, the squeezing term $\xi_n$ denotes the contribution to DCE by the variation of eigenfrequencies, while the acceleration term $\mu_{mn}^{\pm}$ denotes the contribution by the variation of eigenfunctions.

Considering a static system with perturbation parameter $\delta_t = 0$ when $t<0$, we define the Fock states of vacuum and the initial creation and annihilation operators, which satisfies:

$$a_n^{(0)}\left|0\right\rangle = 0. \quad \text{(S3.13)}$$

Considering a permittivity $\varepsilon(t)$ that dynamically varies in the duration between 0 and $t_f$ and is at rest for both $t<0$ and $t>t_f$, the initial and final state operators are linked through a Bogoliubov transformation:

$$a_n\left(t_f\right) = \sum_m \left[ A_{mn}\left(t_f\right)e^{i\omega_{m0}t_f}a_m^{(0)} + B_{mn}^*\left(t_f\right)e^{-i\omega_{m0}t_f}a_m^{+(0)} \right]. \quad \text{(S3.14)}$$

The evolution equations of the connected matrixes $A_{mn}(t)$ and $B_{mn}(t)$ can be obtained from Eq. (S3.11) with initial values $A_{mn}(t\leq 0) = \delta_{mn}$ and $B_{mn}(t\leq 0) = 0$:

$$\begin{cases} \dfrac{\partial}{\partial t}A_{mn} = -i\omega_n A_{mn} + \xi_n B_{mn} - \sum_s \left( \mu_{sn}^+ A_{ms} + \mu_{sn}^- B_{ms} \right) \\ \dfrac{\partial}{\partial t}B_{mn} = +i\omega_n B_{mn} + \xi_n A_{mn} - \sum_s \left( \mu_{sn}^+ B_{ms} + \mu_{sn}^- A_{ms} \right) \end{cases}. \quad \text{(S3.15)}$$

Eq. (S3.15) is solved by using a five-stage Runge–Kutta numerical method. For time $t>t_f$, the final number of excited photons with energy $\omega_n(t_f)$ presented in the initial vacuum is:

$$N_n\left(t_f\right)=\left\langle 0\left|a_n^+\left(t_f\right)a_n\left(t_f\right)\right|0\right\rangle=\sum_m\left|B_{mn}\left(t_f\right)\right|^2 \tag{S3.16}$$

The expectation value is also calculated at different time between 0 and $t_f$. By doing so, $t_f$ is interpreted as a continuous variable such that Eq. (S3.16) becomes a continuous function of time; i.e. $N_n\left(t_f\right)\to N_n\left(t\right)$.

To be consistent with Eq. (5) in the main manuscript, we again use a transformation as follows:

$$\overline{A}_{mn}=A_{mn}\exp\left(+i\omega_{n0}t\right),\overline{B}_{mn}=B_{mn}\exp\left(-i\omega_{n0}t\right), \tag{S3.17}$$

Then we induce a new set of linear equations:

$$\begin{cases}\frac{\partial}{\partial t}\overline{A}_{mn}=-i\left(\omega_n-\omega_{n0}\right)\overline{A}_{mn}+\xi_n\overline{B}_{mn}\exp\left(+2i\omega_{n0}t\right)-\sum_s\left\{\mu_{sn}^+\overline{A}_{ms}\exp\left[+i\left(\omega_{n0}-\omega_{s0}\right)t\right]+\mu_{sn}^-\overline{B}_{ms}\exp\left[+i\left(\omega_{n0}+\omega_{s0}\right)t\right]\right\}\\ \frac{\partial}{\partial t}\overline{B}_{mn}=+i\left(\omega_n-\omega_{n0}\right)\overline{B}_{mn}+\xi_n\overline{A}_{mn}\exp\left(-2i\omega_{n0}t\right)-\sum_s\left\{\mu_{sn}^+\overline{B}_{ms}\exp\left[-i\left(\omega_{n0}-\omega_{s0}\right)t\right]+\mu_{sn}^-\overline{A}_{ms}\exp\left[-i\left(\omega_{n0}+\omega_{s0}\right)t\right]\right\}\end{cases}$$

(S3.18)

in which parameters $\omega_n$, $\xi_n$ and $\mu^{\pm}{}_{sn}$ are all time-dependent.

For a system without dynamically modulated permittivity $\varepsilon(t)$, it is straightforward to obtain a solution with constant $\overline{A}_{mn}$ and $\overline{B}_{mn}$. Here we assume that in the solution of Eq. (S3.17) both matrices are varying slowly with time in a small perturbation configuration. By neglecting the second derivatives (mainly from the time-dependent frequency in $\mu^{\pm}{}_{sn}$) and averaging over those fast oscillations, we obtain:

$$\begin{cases}\frac{\partial}{\partial\tau}\overline{A}_{mn}=\overline{B}_{mn}\cdot\frac{1}{4}w_n\cdot\delta\left(2\omega_{n0}-\omega_d\right)-\sum_s\left[\overline{A}_{ms}\cdot T_{sn}^+\cdot\delta\left(\left|\omega_{n0}-\omega_{s0}\right|-\omega_d\right)+\overline{B}_{ms}\cdot T_{sn}^-\cdot\delta\left(\omega_{n0}+\omega_{s0}-\omega_d\right)\right]\\ \frac{\partial}{\partial\tau}\overline{B}_{mn}=\overline{A}_{mn}\cdot\frac{1}{4}w_n\cdot\delta\left(2\omega_{n0}-\omega_d\right)-\sum_s\left[\overline{B}_{ms}\cdot T_{sn}^+\cdot\delta\left(\left|\omega_{n0}-\omega_{s0}\right|-\omega_d\right)+\overline{A}_{ms}\cdot T_{sn}^-\cdot\delta\left(\omega_{n0}+\omega_{s0}-\omega_d\right)\right]\end{cases},$$

(S3.19)

in which $\tau=\delta_0\omega_d t$ is the normalized time and $T_{mn}^{\pm}=\frac{1}{4}\cdot\left(\sqrt{n/m}\pm\sqrt{m/n}\right)\cdot C_{mn}$ is the effective amplitude of the acceleration effect. Therefore, we can rewrite the equations

in a matrix form:

$$\frac{d}{d\tau}\psi(\tau)=\psi(\tau)\cdot W\,, \tag{S3.20}$$

with $\psi(\tau)=\left(\bar{A}(\tau),\bar{B}(\tau)\right)^{T}$. $W$ is a time-independent matrix and can be expressed as a block matrix form $W$=[$W_{11}$, $W_{12}$; $W_{21}$, $W_{22}$], with:

$$\begin{cases} W_{11,mn}=W_{22,mn}=-T_{mn}^{+}\cdot\delta\left(\left|\omega_{n0}-\omega_{m0}\right|-\omega_{d}\right) \\ W_{12,mn}=W_{21,mn}=\dfrac{1}{4}w_{n}\cdot\delta\left(2\omega_{n0}-\omega_{d}\right)-T_{mn}^{-}\cdot\delta\left(\omega_{n0}+\omega_{m0}-\omega_{d}\right) \end{cases} \tag{S3.21}$$

This is a standard multivariate linear differential equation, and we can solve it directly. With matrix $W$ diagonalized as $W\cdot V=V\cdot\Lambda$, the solution is expressed as:

$$\psi(\tau)=\psi(0)\cdot V\cdot\exp(\Lambda\tau)\cdot V^{-1}. \tag{S3.22}$$

Then by using Eq. (S3.16), we can obtain the spectra of excited photons at different time.

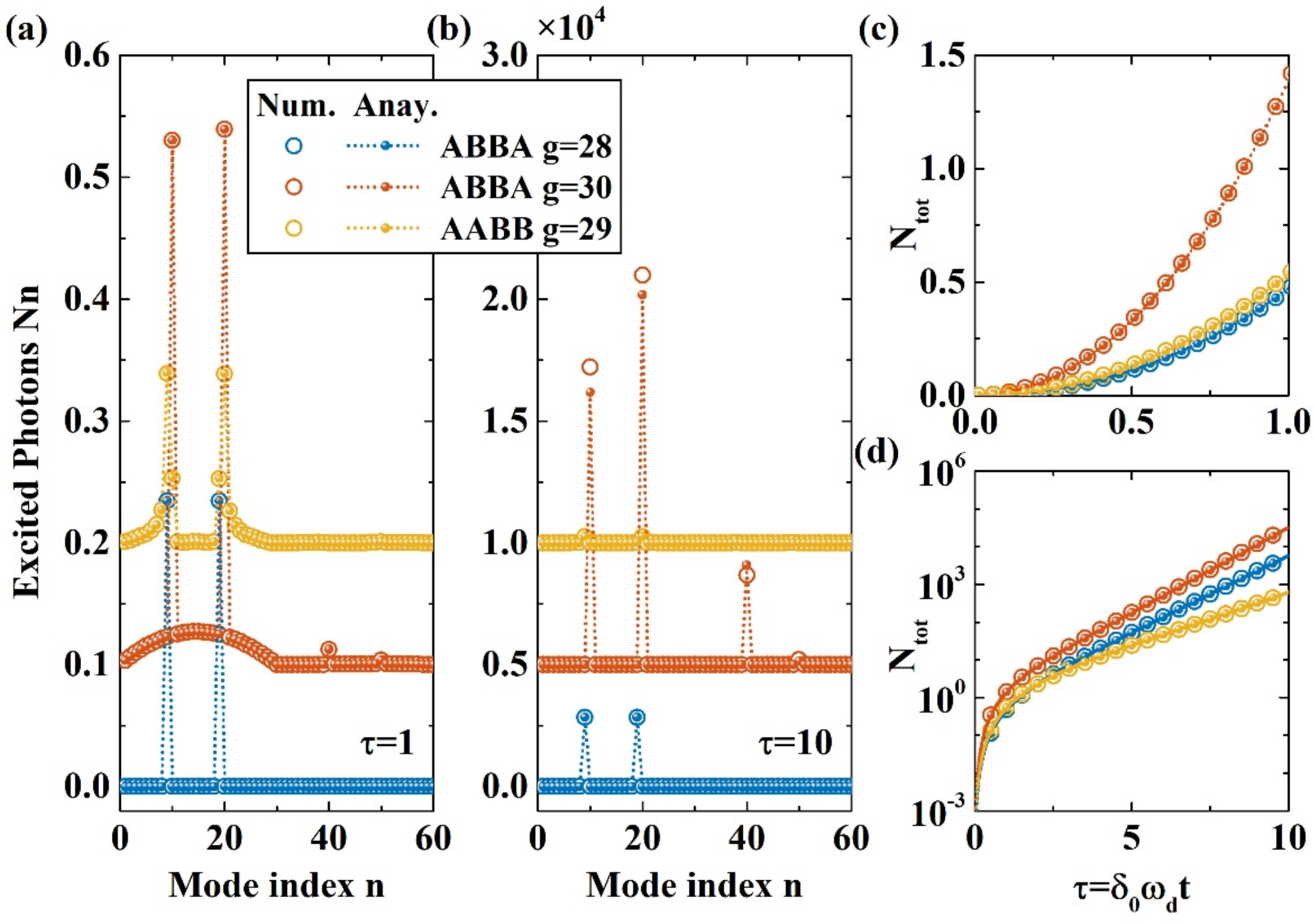

Figure S2. Schematics of the excitation photon spectra for three typical cases at (a) $\tau$=1 and (b) $\tau$=10. To be clearer, we shift different curves along $y$ direction by 0.1 and 5000, respectively. (c) and (d) show the evolution of total photon numbers in short/long time. The results obtained by numerical method are indicated by circles,

and those obtained by analytical method are indicated by dotted line and balls. In all of these calculations, we choose $N_c$= 5 and $\delta_0$= 0.001.

To prove the accuracy of the approximate analytical solution in Eq. (S3.22), we compare it with the numerical method (Solving Eq. (S3.15) by five-stage Runge–Kutta numerical method) for three typical photon spectra results and the evolution of total photon numbers, as shown in Fig. S2. Obviously, they almost perfectly match with each other.

However, for the case with $g$=30 in ABBA system, we find the analytical solution has a small shift at time $\tau$=10, shown in Fig. S2(b). This deviation comes from the second derivative of $\mu^{\pm}$ inside $\sqrt{\omega_n/\omega_m}$ term and its reciprocal, which is time-dependent. For simplification, we have ignored the small time-variation of $\sqrt{\omega_n/\omega_m}$ and directly replace it by $\sqrt{\omega_{n0}/\omega_{m0}} = \sqrt{n/m}$. It turns out that this small deviation is only significant after a long period of time. Moreover, it only occurs in a symmetric system with $g=2n_K$, where the photon modes with significant frequency shift at the Brillouin zone boundary are excited.

The excited spectra in a number of different cases have been discussed in the main manuscript. Here, we focus on the excitation of photons $N_n$ with $n$=40 and $n$=50 in $g$=30 case, which comes from the scattering effects in $W_{11}$ and $W_{22}$ terms. They are responsible for the small differences between those excitations with $n$=10 and $n$=20, due to different scattering efficiencies. Besides, only the squeezing term $w_n$ contributes to the photons with $n$=15 in the $g$=30 case, while there is no contribution from the acceleration term because $C_{nn}$ is always zero due to its anti-symmetric property $C_{mn}$ = -$C_{nm}$.

For all the three resonant excitation conditions shown in Fig. S2(a-b), the total number of created photons $N_{tot}$ (sum over excited photons in all modes) exhibits a quadratic growth in a short time scale, while an exponential growth in a long time scale, as shown in Fig. S2(c-d). This evolution pattern is consistent with the law $N_{tot}(t) = \sinh^2(\beta \cdot \delta_0 \omega_d t)$ in the homogeneous cavities, with $\beta$ being a dimensionless constant to describe the rate of DCE Photon generation process that depends on the specific configuration of the system. The consistency of these evolution patterns also implies that the photon pairs are created through a DCE process.

Indeed, the exponential term in the analytical solution Eq. (S3.22) is directly responsible for this different growth pattern in the short/long time scale. From Eq. (S3.16) the excited photon numbers are related to the square of $B_{mn}$ matrix elements. From the Taylor expansion of an exponential function, it is obvious that the total number of created photons exhibits a quadratic growth in a short-time scale, while an exponential development in a long-time scale, exactly like that exhibited in Fig. S2(c-d).

# 4. Results in an asymmetric PhC system

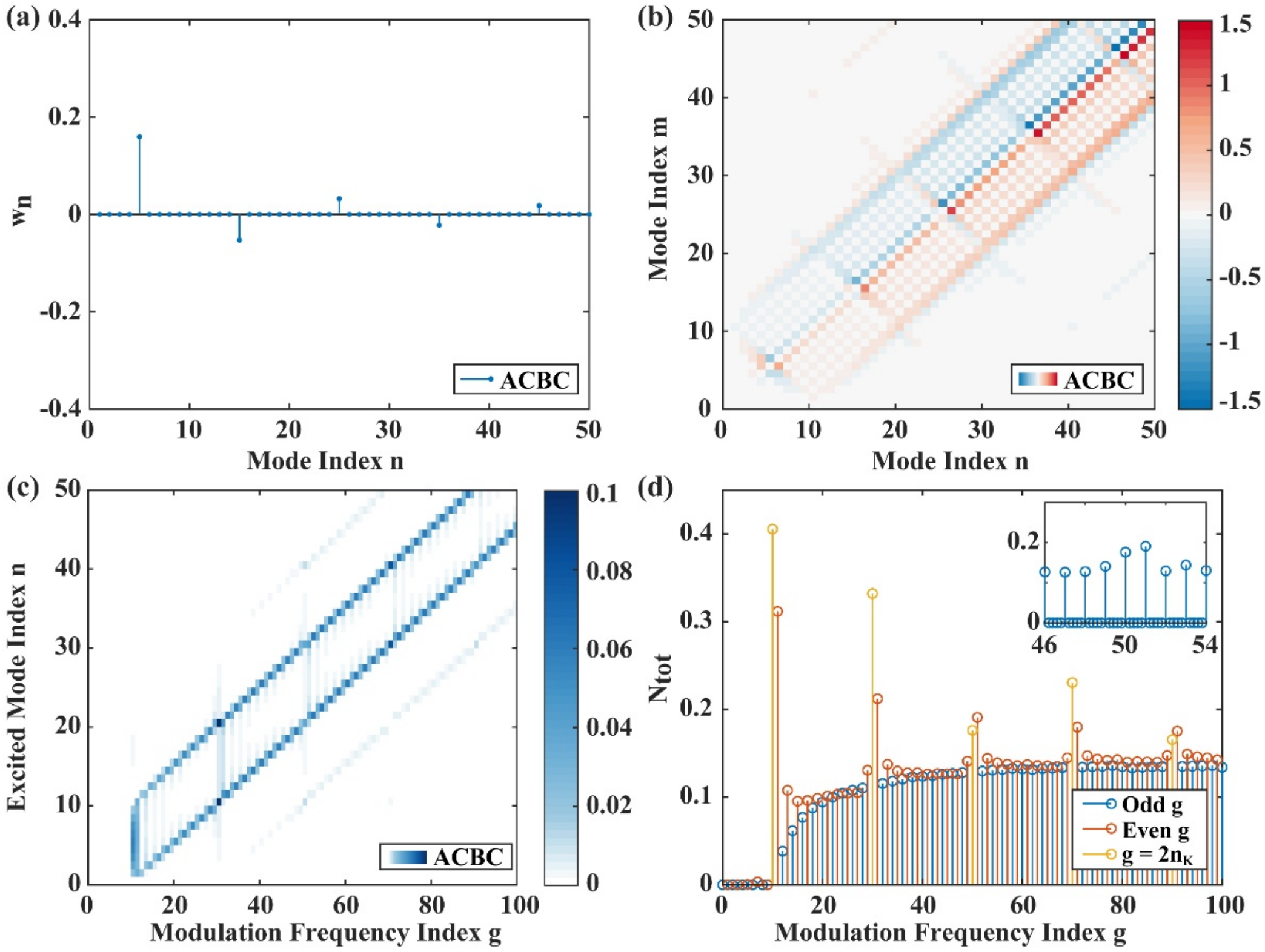

Figure S3 (a-b) Schematics of the amplitude function in (a) squeezing and (b) acceleration terms for each eigenmode in ACBC system (c) Schematics of the created photon spectra $N_n$ with different modulation frequency $\omega_d$ at $\tau$=1 in ACBC system. (d) shows the total number of created photons for even and odd $g$ at the same time. Insets show the results for arbitrary $g$ parameter. In all of these calculations, we choose $N_c = 5$ and solve them through the analytical method.

Considering the DCE process in a PhC system with a general perturbation that is neither symmetric nor anti-symmetric, such as the distribution "ACBC", with "C" a slab with a constant permittivity $\varepsilon_C(t) = \varepsilon_0$.

The asymmetric geometry imposes less restrictions on either the squeezing or the acceleration terms. In this ACBC configuration, the squeezing term $w_n$ is nonzero when $n= n_K$, as shown in Fig. S3(a). Thus, for the mode pairs satisfying $n\pm m= 2l,$ the acceleration term $C_{mn}$ is nonzero when $l=n_K$, which is the same as in the ABBA configuration. Besides, for the mode pairs satisfying $n\pm m= 2l+1,$ because the distribution does not have a well-defined symmetry, $C_{mn}$ that connects two eigenmodes with different parities are nonzero, which is the same as that in the AABB configuration, , as shown in Fig. S3 (b). Thus, in the ACBC configuration, both the modulation frequencies with even and odd integer $g$ parameters can resonantly excite the system, as shown in Fig. S3(c-d).

# 5. Input-Output Theory

Here we calculate the radiation from a PhC cavity, in which one PEC boundary is replaced by a high reflectivity mirror. We use $Q$-factor of the first K-mode to describe the coupling between the mode and the free space radiation, with the coupling strength given by:

$$\kappa_K = \frac{\omega_K}{2Q_K} = \frac{c\pi/D}{2Q_K}, \tag{S5.1}$$

in which $\omega_K = c\pi/D$ (D is the periodicity) is independent of the repeated number of unitcells $N_c$. In the following analysis, all the modes have nearly the same coupling strength as the K-mode. i.e. $\kappa = \kappa_K$. By employing the standard input-output theory, and allowing an external field $a_{in}$ to be coupled into the cavity, the time evolution of creation operator for the $n$-th eigenmode is given by:

$$\frac{\partial \hat{a}_n}{\partial t} = -i\omega_n \hat{a}_n + \xi_n \hat{a}_n^+ - \sum_m \left[\mu_{mn}^+ \hat{a}_m + \mu_{mn}^- \hat{a}_m^+\right] - \kappa \hat{a}_n - \sqrt{2\kappa}\hat{a}_{in}, \tag{S5.2}$$

where all the parameters are defined the same as that in Eq. (S3.11). Similarly, transforming to the interaction operator with $\bar{a}_n^{\pm} = a_n^{\pm} \cdot \exp\left( i\omega_{n,0} t\right)$, we have:

$$\frac{\partial \bar{a}_n}{\partial t} = -i\left(\omega_n - \omega_{n,0}\right)\bar{a}_n + \xi_n \bar{a}_n^+ e^{2i\omega_{n0}t} - \sum_m \begin{bmatrix} \mu_{mn}^+ \bar{a}_m e^{-i(\omega_{m0}-\omega_{n0})t} \\ +\mu_{mn}^- \hat{a}_m^+ e^{i(\omega_{m0}+\omega_{n0})t} \end{bmatrix} - \kappa \bar{a}_n - \sqrt{2\kappa}\hat{a}_{in} e^{i\omega_{n0}t}, \tag{S5.3}$$

Averaging over those fast oscillations, we have:

$$\frac{\partial \bar{a}_n}{\partial t} = \frac{1}{4} w_n \delta_0 \omega_d \delta\left(2\omega_{n,0} - \omega_d\right)\cdot \bar{a}_n^+ - \delta_0 \omega_d \sum_m \begin{bmatrix} T_{mn}^+ \delta\left(\left|\omega_{n,0} - \omega_{m,0}\right| - \omega_d\right)\cdot \bar{a}_m \\ +T_{mn}^- \delta\left(\omega_{n,0} + \omega_{m,0} - \omega_d\right)\cdot \hat{a}_m^+ \end{bmatrix} - \kappa \bar{a}_n - \sqrt{2\kappa}\hat{a}_{in} e^{i\omega_{n0}t}, \tag{S5.4}$$

with all the parameters defined the same as that in Eq. (S3.19).

With out-coupling of the photons into free space, the energy of the internal mode does not keep growing over time as that in a closed system. We can solve its steady state through a harmonic calculation in the frequency domain.

Following the convention of Clerk et.al [5] for the Fourier transform of time dependent operators, we have $a(\omega)=\int_{-\infty}^{\infty} dt \cdot a(t)\exp(i\omega t)$ and $a^+(\omega)=\left[a(-\omega)\right]^+ =\int_{-\infty}^{\infty} dt \cdot a^+(t)\exp(i\omega t)$. In the interaction representation $\bar{a}_n^{\pm} = a_n^{\pm} \cdot \exp\left( i\omega_{n,0} t\right)$ implies that the *n*-th mode operator at a certain frequency appears shifted in the rotating frame by $\omega_{n,0}$, with $\bar{a}_n(\omega) = a_n(\omega + \omega_{n,0})$ and $\bar{a}_n^+(\omega) = \left[a_n(\omega_{n,0} - \omega)\right]^+$.

For the Fourier transformed mode operators, we can immediately obtain the frequency domain equation by inserting the Fourier transform equations into Eq. (S5.4), multiplying it by exp(*ivt*) and integrating over time *t*:

$$\begin{aligned} -iv\bar{a}_n(v) &= \frac{1}{4} w_n \delta_0 \omega_d \delta\left(2\omega_{n,0} - \omega_d\right)\cdot \bar{a}_n^+(v) \\ &\quad -\delta_0 \omega_d \sum_m \begin{bmatrix} T_{mn}^+ \delta\left(\left|\omega_{n,0} - \omega_{m,0}\right| - \omega_d\right)\cdot \bar{a}_m(v) \\ +T_{mn}^- \delta\left(\omega_{n,0} + \omega_{m,0} - \omega_d\right)\cdot \hat{a}_m^+(v) \end{bmatrix} - \kappa \bar{a}_n(v) - \sqrt{2\kappa}\hat{a}_{in}\left(v + \omega_{n0}\right), \end{aligned} \tag{S5.5}$$

Transforming the equations back into the nonrotating frame, and replacing the frequency parameter *v* by $\omega - \omega_{n,0}$, we obtain the following evolution equation for

$a(\omega)$:

$$-i\left(\omega-\omega_{n0}\right)a_n\left(\omega\right)=\frac{1}{4}w_n\delta_0\omega_d\delta\left(2\omega_{n,0}-\omega_d\right)\cdot\left[a_n\left(2\omega_{n,0}-\omega\right)\right]^+$$
$$-\delta_0\omega_d\sum_m\begin{bmatrix}T_{mn}^+\delta\left(\left|\omega_{n,0}-\omega_{m,0}\right|-\omega_d\right)\cdot a_m\left(\omega-\omega_{n,0}+\omega_{m,0}\right)\\+T_{mn}^-\delta\left(\omega_{n,0}+\omega_{m,0}-\omega_d\right)\cdot\left[a_m\left(\omega_{n,0}+\omega_{m,0}-\omega\right)\right]^+\end{bmatrix}-\kappa a_n\left(\omega\right)-\sqrt{2\kappa}\hat{a}_{in}\left(\omega\right)$$

, (S5.6)

which can be further written as:

$$-i\left(\omega-\omega_{n0}\right)a_n\left(\omega\right)=\frac{1}{4}w_n\delta_0\omega_d\delta\left(2\omega_{n,0}-\omega_d\right)\cdot\left[a_n\left(\omega_d-\omega\right)\right]^+$$
$$-\delta_0\omega_d\sum_m\begin{bmatrix}T_{mn}^+\delta\left(\omega_{n,0}-\omega_{m,0}-\omega_d\right)\cdot a_m\left(\omega-\omega_d\right)\\+T_{mn}^+\delta\left(\omega_{m,0}-\omega_{n,0}-\omega_d\right)\cdot a_m\left(\omega+\omega_d\right)\\+T_{mn}^-\delta\left(\omega_{n,0}+\omega_{m,0}-\omega_d\right)\cdot\left[a_m\left(\omega_d-\omega\right)\right]^+\end{bmatrix}-\kappa a_n\left(\omega\right)-\sqrt{2\kappa}\hat{a}_{in}\left(\omega\right),$$

(S5.7)

Clearly, only the creation/annihilation operators at frequencies $\omega_d-\omega$, $\omega$ and $\omega\pm\omega_d$ can couple to each other. Further, we can ignore the scattering effect by $T_{mn}^+$ and $a_m\left(\omega\pm\omega_d\right)$ in the frequency region ranging from 0 to $\omega_d$, due to the small occupation number of photons at high frequency region under low temperatures.

In this way, the evolution equations for the operators $a_n\left(\omega\right)$ is:

$$\left[\chi_n\left(\omega\right)\right]^{-1}a_n\left(\omega\right)-F_n^I\left(\omega_d\right)\cdot\left[a_n\left(\omega_d-\omega\right)\right]^++\sum_m F_{mn}^{II}\left(\omega_d\right)\cdot\left[a_m\left(\omega_d-\omega\right)\right]^+=-\sqrt{2\kappa}\hat{a}_{in}\left(\omega\right).$$

(S5.8)

In the above expression, we have defined $\chi_n\left(\omega\right)=\left[\kappa-i\left(\omega-\omega_{n0}\right)\right]^{-1}$ as the electric cavity susceptibility, $F_n^I\left(\omega_d\right)=\frac{1}{4}w_n\delta_0\omega_d\delta\left(2\omega_{n,0}-\omega_d\right)=\left(F_n^I\right)^*$ as the squeezing contribution, and $F_{mn}^{II}\left(\omega_d\right)=\delta_0\omega_d T_{mn}^-\delta\left(\omega_{n,0}+\omega_{m,0}-\omega_d\right)=\left[F_{mn}^{II}\left(\omega_d\right)\right]^+$ as one of the acceleration contribution.

Similarly, we can obtain the evolution equation for operators $a_n\left(\omega_d-\omega\right)$ by changing the frequency in Eq. (S5.8) from $\omega$ to $\omega_d$ -$\omega$, and obtain the evolution equation for operators $\left[a_n\left(\omega\right)\right]^+$ and $\left[a_n\left(\omega_d-\omega\right)\right]^+$ through conjugation. They construct a

closed set of equations:

$$
\begin{cases}
\left[\chi_n(\omega)\right]^{-1} a_n(\omega) - F_n^{I}(\omega_d)\cdot\left[a_n(\omega_d-\omega)\right]^{+} + \sum\limits_m F_{mn}^{II}(\omega_d)\cdot\left[a_m(\omega_d-\omega)\right]^{+} = -\sqrt{2\kappa}\hat{a}_{in}(\omega) \\
\left[\chi_n(\omega_d-\omega)\right]^{-1} a_n(\omega_d-\omega) - F_n^{I}(\omega_d)\cdot\left[a_n(\omega)\right]^{+} + \sum\limits_m F_{mn}^{II}(\omega_d)\cdot\left[a_m(\omega)\right]^{+} = -\sqrt{2\kappa}\hat{a}_{in}(\omega_d-\omega) \\
\left[\chi_n^{*}(\omega)\right]^{-1}\left[a_n(\omega)\right]^{+} - F_n^{I}(\omega_d)\cdot a_n(\omega_d-\omega) + \sum\limits_m F_{mn}^{II}(\omega_d)\cdot a_m(\omega_d-\omega) = -\sqrt{2\kappa}\hat{a}_{in}^{+}(\omega) \\
\left[\chi_n^{*}(\omega_d-\omega)\right]^{-1}\left[a_n(\omega_d-\omega)\right]^{+} - F_n^{I}(\omega_d)\cdot a_n(\omega) + \sum\limits_m F_{mn}^{II}(\omega_d)\cdot a_m(\omega) = -\sqrt{2\kappa}\hat{a}_{in}^{+}(\omega_d-\omega)
\end{cases}
\tag{S5.9}
$$

Together with the equation for the output modes, $\hat{a}_{out}(\omega) = \hat{a}_{in}(\omega) + \sum_n \sqrt{2\kappa}\cdot a_n(\omega)$, we can solve the relationship between the output photon operators $\left[\hat{a}_{out}(\omega)\right]^{\pm}$ and the input photon operators $\left[\hat{a}_{in}(\omega)\right]^{\pm}$ and $\left[\hat{a}_{in}(\omega_d-\omega)\right]^{\pm}$.

This relation can be written in the general form:

$$\hat{a}_{out}(\omega) = c_{+}\cdot\hat{a}_{in}(\omega) + c_{-}\cdot\hat{a}_{in}(\omega_d-\omega) + d_{+}\cdot\left[\hat{a}_{in}(\omega)\right]^{+} + d_{-}\cdot\left[\hat{a}_{in}(\omega_d-\omega)\right]^{+}, \tag{S5.10}$$

then the number of output photons satisfies:

$$n_{out}(\omega) = \left(|c_{+}|^2 + |d_{+}|^2\right)\cdot n_{in}(\omega) + \left(|c_{-}|^2 + |d_{-}|^2\right)\cdot n_{in}(\omega_d-\omega) + \left(|d_{+}|^2 + |d_{-}|^2\right), \tag{S5.11}$$

with the number of input photons given by $n_{in}(\omega) = 1/\left[\exp(\ \omega/k_B T) - 1\right]$, which represents the photon numbers induced by classical thermal fluctuations at a certain temperature $T$ ($\hbar$ is the reduced Planck Constant and $k_B$ is the Boltzmann constant). The first two terms are of classical origin and the last term is a purely quantum mechanical effect which originates from the vacuum quantum fluctuations, or in other words, from the DCE. We note that the vacuum fluctuation dominates the output flux with $n_{in}(\omega) \approx 0$ at low temperature.

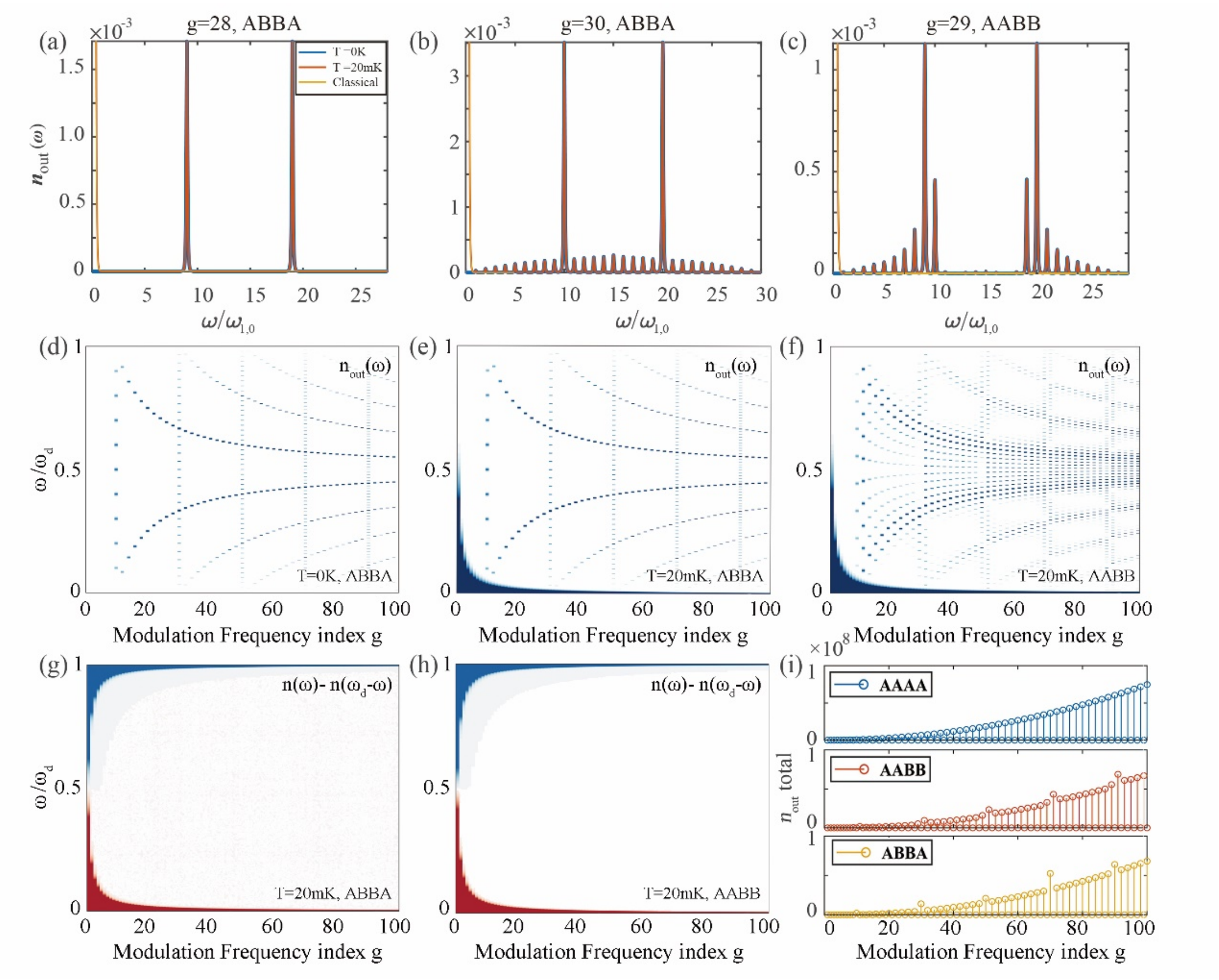

Figure S4. Schematics of the output photons spectra from a dynamic PhC cavity for three typical cases with (a) $g$=28 and (b) $g$=30 in ABBA system and (c) g=29 in AABB system. (d-f) show the evolution of output photons spectra in two systems with various modulation frequency at temperature (d) $T=0K$ and (e-f) $T=20mK$. (g-f) show the difference between output spectra of two correlated frequencies $\omega$ and $\omega_d$ -$\omega$ in two system at temperature 20mK. (i) shows the total number of created photons of three PhC systems with various modulation frequency at temperature T=0K. In all of these calculations, we choose $N_c = 5$, $d = 1.0mm$, $\delta_0 = 0.001$, $\varepsilon_0 = \mu_0$ =1 and $Q_K = 50$.

For a general case with the modulation frequency parameter $g \neq 2s$ in the ABBA configuration and $g \neq 2s+1$ in the AABB configuration ($s$ is an arbitrary integer number), both contributions $F_n^{I}(\omega_d)$ and $F_n^{II}(\omega_d)$ are zero. In this way, we can solve the equations with $\hat{a}_{out}(\omega) = \left[1 - \sum_n 2\kappa\chi_n(\omega)\right] \cdot \hat{a}_{in}(\omega)$, which represents the resonance enhancement for classical thermal fluctuations, but without any contribution from DCE.

For DCE resonance cases with the typical $g$ parameters discussed in the main text, while we cannot find a simple analytical solution for the evolution equations, we can solve them using the numerical method. Next, we will show the results for some typical cases, which has output photon flux patterns that are closely related to the photon distributions in the corresponding closed systems.

Choosing $Q_K$ =50, we calculate the output photons spectra, as shown in Fig. S4. The output spectra of the three typical modulation cases shown in Fig. S4(a-c) are strongly related to the excited photon spectra in the corresponding closed PhC cavity shown in Fig. S2. This is because the output spectra are radiated by the excited photons inside the cavity. More calculations for output spectra with various modulation frequencies, as shown in Fig. S4 (d-f), verify these direct connections. However, the output spectra exhibit a certain bandwidth, whereas the modes in the closed system are infinitely sharp. This is a general difference between the closed and open system.

Besides, at a low temperature $T$ =20$mK$ with $k_B T << \omega_d$, we can safely ignore the effects of thermal excitation because the spectrum is exactly the same as that at $T$ = 0K, except for close to zero frequency where thermal fluctuations are diverging. More importantly, the photon spectra of the two correlated frequencies $\omega$ and $\omega_d$ -$\omega$ are identical to each other, except for at low frequencies where strong thermal fluctuations are present, as shown in Fig. S4(a-c) and Fig. S4(g-h). This serves as an evidence that at low temperature the radiation originates from vacuum quantum fluctuations rather than classical thermal fluctuations. Moreover, we study the total DCE photons flux $n_{\text{tot}}$ at temperature $T$=0$K$ (to avoid the divergence of classical thermal fluctuation near zero frequency) defined by $n_{tot} = 1/2\pi \cdot \int_0^{\omega_d} d\omega \cdot n_{out}(\omega)$, shown in Fig. S4(i). We find an enhancement at the modulation frequency with $g$ =2$n_K$ in the ABBA configuration, and $g$ =2$n_K$±1 in the AABB configuration, which is similar to the closed system. The enhancement factor is not very large here due to a small number of unitcells $N_c$ =5, but it can be further enhanced with a larger $N_c$.

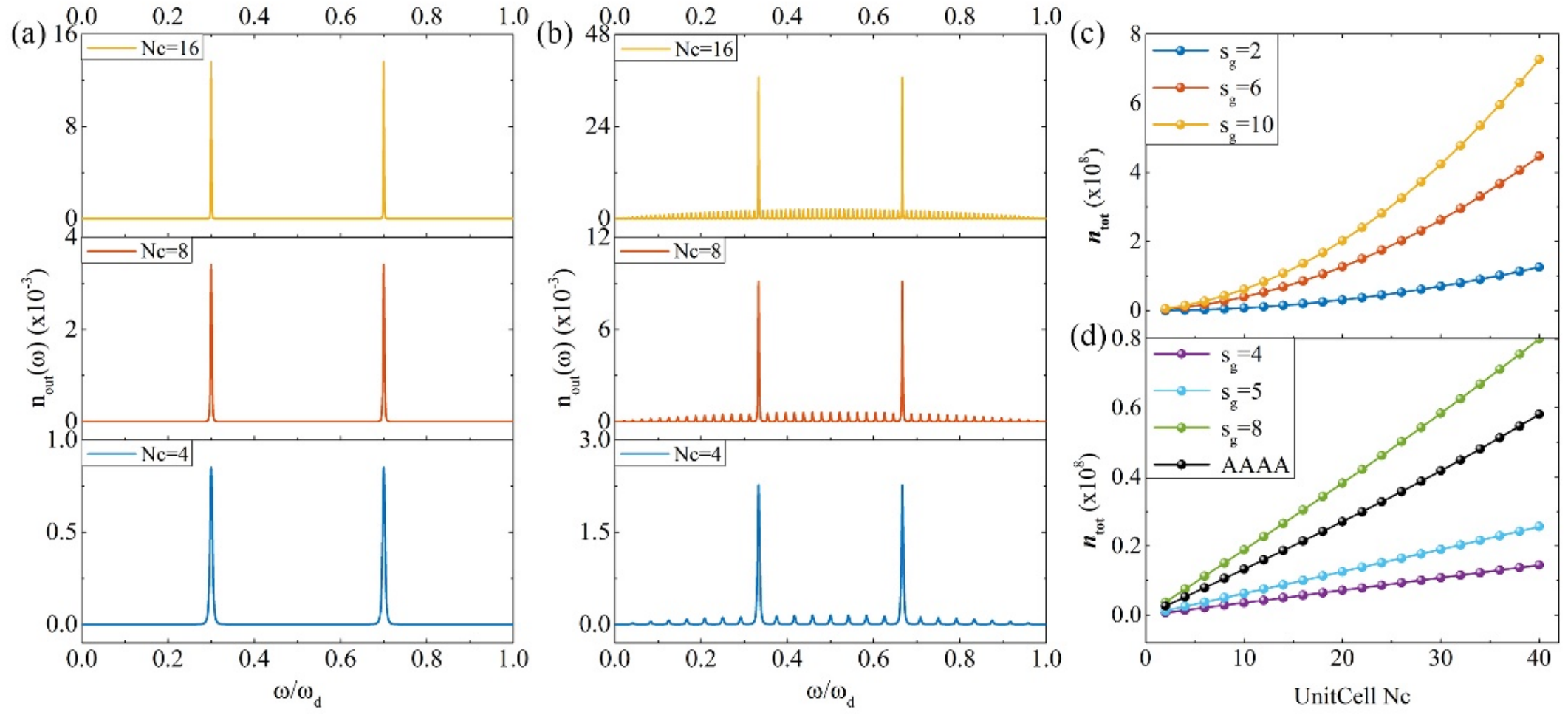

Figure S5 (a-b) Schematics of the output photons spectra $n_{\mathrm{out}}(\omega)$ in the ABBA PhC configuration, with same boundary condition but different repeated number of unit cells $N_c$ at $T$= 0K, with modulation frequency (a) $\omega_d=5N_c\cdot\pi c/L$ ($s_g$=5) and (b) $\omega_d=6N_c\cdot\pi c/L$ ($s_g$=6) respectively. (c-d) show total photons flux $n_{tot}$ in the ABBA PhC cavity with different $s_g$ parameter, and in the AAAA system with $s_g$ =6, with same boundary condition but different $N_c$. In all of these calculations, we choose $Q_K$=10$N_c$, $d$ = 1.0$mm$, $\delta_0$ = 0.001, $\varepsilon_0 = \mu_0$=1 and temperature $T$ =0$K$.

To discuss the effect of the number of unit cell $N_c$, we calculate the output spectra $n_{\mathrm{out}}(\omega)$ in the ABBA PhC system with different $N_c$ and $s_g$ at $T$ =0$K$, as shown in Fig. S5(a-b). As we know, the decay rate of a homogeneous FP cavity should be inversely proportional to the time for photons propagating in a round trip between two boundaries. Thus, under the same boundary condition, $Q_K$ is proportional to the length of our PhC cavity, i.e. $Q_K \propto N_c$.

Choosing $Q_K$=10$N_c$, the output spectra with different $N_c$ are shown in Fig. S5. For a typical excitation with $s_g$=5, a larger $N_c$ and $Q_K$ factor induce a higher peak which is proportional to the square of the $Q$-factor, and a narrower resonance line width ($\kappa$) which is inversely proportional to the $Q$-factor, as shown in Fig. S5(a). As a result, the enhancement factor for the total DCE photons flux $n_{\mathrm{tot}}$ is linearly scaled with $N_c$, as shown in Fig. S5(d) with $s_g$ =4, 5, 8. Similar results occur in the AAAA system.

However, for a typical excitation with $s_g$=6 ($s_g$ =2$n_K$), except for those properties shown in $s_g$=5 case, there is an additional enhancement arising from the increasing density of cavity modes with an increasing $N_c$, as shown in Fig. S5(b). This induces an extra enhancement for the total DCE photons flux $n_{\mathrm{tot}}$, which is quadratically related to $N_c$, as shown in Fig. S5(c) with $s_g$=2, 6, 10. In a real SQUID circuits, the $Q$-factor can reach about 10000 [6], which means that a strong enhancement can be achieved for those modulations with $g = 2n_{\mathrm{K}}$ in a system with very large $N_c$.

# 6. SQUID circuits Design

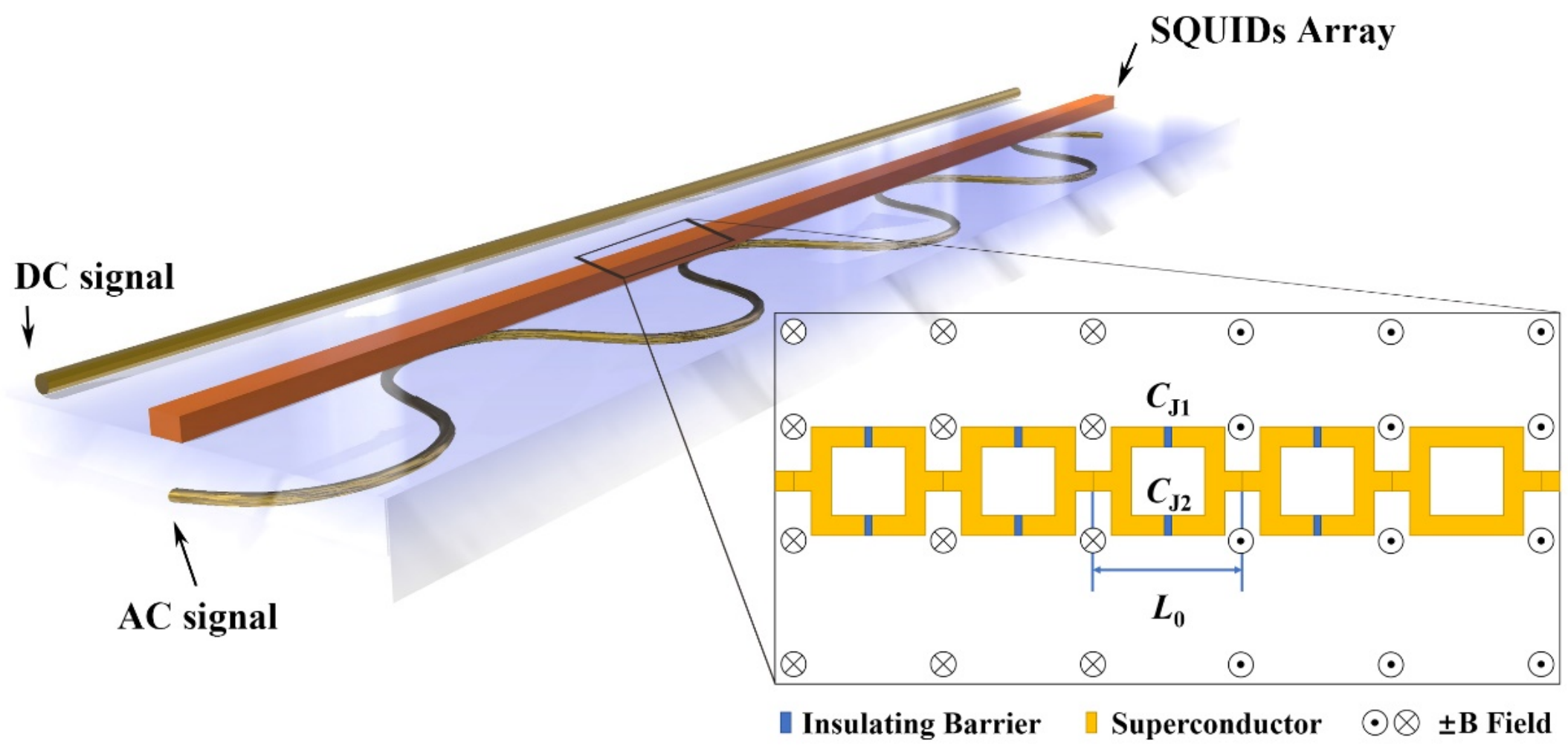

Figure S6. Schematics of the SQUID circuit Design.

In this section, we will use the typical parameters of SQUID electric circuit provided in Ref [6,7] to design a realistic SQUID based 1D PhC system. The two junctions in SQUID with physical size about $L_0$ = 450nm have equal capacitances $C_{J1}$= $C_{J2}$ =$C$/2 with C = 90fF. With an external flux $f$, the SQUID can be considered as a tunable Josephson inductance, as long as the SQUID is weakly excited and the dynamic modulation frequency is small compared to the plasma frequency $\omega_p(f) = 1/\sqrt{CL_J(f)}$. For a given $\omega_p/2\pi = 36GHz$ in Ref. [6,7], we can estimate its effective conductance $L_{J0} = L_J(f_0) = 2.1717\times10^{-10}H$ with a static external flux $f_0$.

For a frequency region far below the plasma frequency, we can ignore its dispersion and only consider the dependence of the effective conductance of SQUID circuit on the magnetic flux.

We consider an on-chip SQUIDs series array embedded in a coplanar transmission line with 1200 SQUIDs units (total length 0.54mm) and use the excitation setup in [6] to map the circuits to our design. Here a DC-signal generates a static external flux, and a AC-signal is used to dynamically tune the external flux and the effective conductance of the SQUID circuit. The only difference is that we need a more complex configuration of AC on-chip flux line – a S-shaped wiggling line consisting of connected half circles of opposite orientations. The different winding directions between neighboring half circles introduce opposite directions of the perturbated magnetic flux, as shown in Fig. S6. This AC signal line can be placed underneath the superconducting circuit layer, separated by a dielectric layer. In our design, each half-circle coil has a diameter $d$ = 2.7um covering 6 SQUID units. In this way, we can implement an ABBA or AABB unit-cell in our main text with a length of $10.8\mu m$, which consists of 24 SQUIDs units and 4 half-circle turns. With these design parameters, the resonance frequency of the first K-mode is $\omega_K = v\pi/D \approx 13.1\% \cdot \omega_p$, and we can ignore the dispersion of the effective conductance with the modulation frequency index $g$ ranging from 0 to about $8N_c$.Thus, the whole sample with 1200 SQUID units can mimic a PhC system with $N_c$ =50. One end of the circuit is grounded to mimic the PEC boundary conditions, while the other end is coupled to a capacitor to mimic a semitransparent mirror for collecting the output DCE photon flux signal.

We use the input-output model established in Sec. 5 to calculate the photon flux in this realistic SQUID-based design. First, we map the effective parameters from [$L_J$, $C$] to [$\varepsilon$, $\mu$] through the wave velocity $v = L_0\big/\sqrt{CL_{J0}} = c\big/\sqrt{\varepsilon_r\mu_r}$ and the impedance $Z = \sqrt{L_{J0}/C} = Z_0\sqrt{\mu_r/\varepsilon_r}$. The effective relative permittivity is about $\varepsilon_r$ =22600 and relative permeability is about $\mu_r$ = 383. Then we assume a dynamic modulation of the effective permeability $\mu_r$ (corresponding to a tunable $L_J$) satisfying $\mu_r(x,t) = \mu_r(1 \pm \delta_t)$. We use the parameters $\delta_0$=0.001, and $Q_K$ =1000 to calculate the radiation photon flux, as shown in Fig. S7.

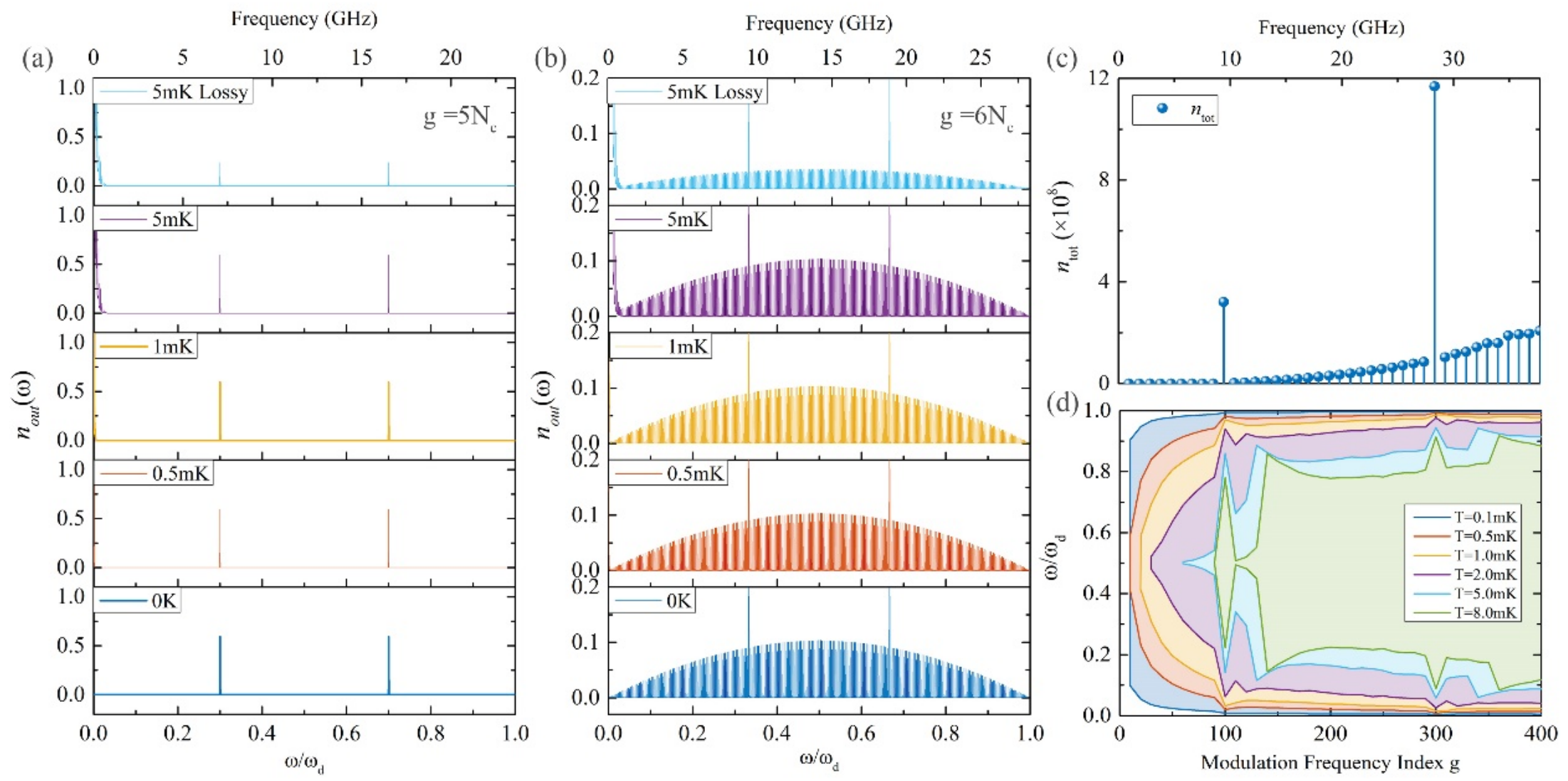

Figure S7. (a-b) Schematics of the output photons spectra $n_{out}(\omega)$ in the realistic SQUID based ABBA PhC system, with the same $Q_K$ =1000, $N_c$ =50 but different temperature and material loss, with modulation frequency (a) $\omega_d=5N_c\cdot\pi c/L$ ($s_g$=5) and (b) $\omega_d=6N_c\cdot\pi c/L$ ($s_g$=6) respectively. (c) shows total photons flux $n_{tot}$ radiated from this cavity with various modulation frequency at 0K. (d) shows the DCE critical line at different temperatures.

Fig. S7 (a-b) show the output spectra at different modulation frequencies, which are in good agreement with the analysis in section 5. Here we excite the resonances with lower frequencies compared with that in Fig. S4, which requires a much lower temperature 2~5mK (realizable through 3He-4He dilution refrigeration technology) to satisfy $k_B T << \ \omega_d$ for minimizing the contribution of the thermal excitations. For g = $2n_K$ with $g$ =100 ($n_{tot}$= 3.2×$10^8$) and $g$ = 300 ($n_{tot}$= 1.2×$10^9$), we observe 1-2 orders of magnitude stronger excitations than the other resonant $g$ values (e.g., $n_{tot}$= 3.1×$10^7$ with g=200). This photon flux is possible to be detected under the existing experimental conditions. Indeed, in the real system, we can further enhance this output photon flux by applying a higher $Q$-factor (10000 for a real SQUIDs circular), a larger $N_c$ value (with a larger SQUIDs series), and a stronger modulation amplification $\delta_0$ (it can be as large as 0.25) [6] in the real experiment.

Moreover, to quantitative measure the effect of thermal excitation, we define a DCE noise ratio with $\eta(\omega,T)=\left[n(\omega,T)-n(\omega_d-\omega,T)\right]/\left[n(\omega,T)+n(\omega_d-\omega,T)\right]$. We plot a boundary contour defined by $\eta(\omega,T)=1\%$ at different temperatures to identify the domain dominated by DCE, as shown by different colors in Fig. S7(d). In the figure, the DCE-domain regions are marked by lighter colors on the right side of each boundary line. Clearly, with a lower temperature or a larger modulation frequency, we can detect the DCE-dominated signal in a broader frequency range. Based on the plot, a temperature about 2~5mK is necessary for detecting the DCE signal and excluding the effect of classical thermal fluctuations.

Finally, we consider the influence of dissipation in the electric permittivity. Here, we consider a dissipation factor $\kappa_a$ in the input-output theory to describe SQUID circuits with a relatively low material loss. This dissipation factor modifies the expression of $\chi_n(\omega)$ as $\chi_n(\omega)=\left[\kappa+\kappa_a-i(\omega-\omega_{n0})\right]^{-1}$ in Eq. (S.5.8). In our numerical simulation, similar to a homogeneous cavity, $\kappa_a$ is set to be proportional to the mode index. We find that the only effect of dissipation is the reduction of the intensity of the output photon flux, as shown in Fig. S7 (a-b) for $\kappa_{a,K}=0.1\cdot\kappa_K$, $\kappa_{a,n}=\kappa_{a,K}\cdot n/N_c$, and T = 5mK.

**Reference**

[1] S. G. Johnson, M. Ibanescu, M. Skorobogatiy, O. Weisberg, J. Joannopoulos, and Y. Fink, Physical review E **65**, 066611 (2002).

[2] S. G. Johnson, M. L. Povinelli, M. Soljacic, A. Karalis, S. Jacobs, and J. D. Joannopoulos, Applied Physics B-Lasers and Optics **81**, 283 (2005).

[3] J. D. Joannopoulos, S. G. Johnson, J. N. Winn, and R. D. Meade, *Photonic crystals: molding the flow of light* (Princeton university press, 2011).

[4] S. A. Fulling, *Aspects of quantum field theory in curved spacetime* (Cambridge

university press, 1989), Vol. 17.

[5] A. A. Clerk, M. H. Devoret, S. M. Girvin, F. Marquardt, and R. J. Schoelkopf, Reviews of Modern Physics **82**, 1155 (2010).

[6] M. Sandberg, C. M. Wilson, F. Persson, T. Bauch, G. Johansson, V. Shumeiko, T. Duty, and P. Delsing, Applied Physics Letters **92** (2008).

[7] J. R. Johansson, G. Johansson, C. M. Wilson, and F. Nori, Phys Rev Lett **103**, 147003 (2009).